\documentstyle[epsfig]{mn}

\newif\ifAMStwofonts

\ifoldfss
  \ifCUPmtlplainloaded \else
    \NewTextAlphabet{textbfit} {cmbxti10} {}
    \NewTextAlphabet{textbfss} {cmssbx10} {}
    \NewMathAlphabet{mathbfit} {cmbxti10} {} 
    \NewMathAlphabet{mathbfss} {cmssbx10} {} 
  \fi
  \ifAMStwofonts
    \ifCUPmtlplainloaded \else
      \NewSymbolFont{upmath} {eurm10}
      \NewSymbolFont{AMSa} {msam10}
      \NewMathSymbol{\upi}     {0}{upmath}{19}
      \NewMathSymbol{\umu}     {0}{upmath}{16}
      \NewMathSymbol{\upartial}{0}{upmath}{40}
      \NewMathSymbol{\leqslant}{3}{AMSa}{36}
      \NewMathSymbol{\geqslant}{3}{AMSa}{3E}

      \let\leq=\leqslant 
      \let\geq=\geqslant \let\ge=\geqslant
    \fi
  \fi
\fi 

\ifnfssone
  \newmathalphabet{\mathit}
  \addtoversion{normal}{\mathit}{cmr}{m}{it}
  \addtoversion{bold}{\mathit}{cmr}{bx}{it}
  \newmathalphabet{\mathbfit} 
  \addtoversion{normal}{\mathbfit}{cmr}{bx}{it}
  \addtoversion{bold}{\mathbfit}{cmr}{bx}{it}
  \newmathalphabet{\mathbfss} 
  \addtoversion{normal}{\mathbfss}{cmss}{bx}{n}
  \addtoversion{bold}{\mathbfss}{cmss}{bx}{n}
  \ifAMStwofonts
    \ifCUPmtlplainloaded \else
      \UseAMStwoboldmath
      \makeatletter
      \new@mathgroup\upmath@group
      \define@mathgroup\mv@normal\upmath@group{eur}{m}{n}
      \define@mathgroup\mv@bold\upmath@group{eur}{b}{n}
      \edef\UPM{\hexnumber\upmath@group}
      \new@mathgroup\amsa@group
      \define@mathgroup\mv@normal\amsa@group{msa}{m}{n}
      \define@mathgroup\mv@bold\amsa@group{msa}{m}{n}
      \edef\AMSa{\hexnumber\amsa@group}
      \makeatother
      \mathchardef\upi="0\UPM19
      \mathchardef\umu="0\UPM16
      \mathchardef\upartial="0\UPM40
      \mathchardef\leqslant="3\AMSa36
      \mathchardef\geqslant="3\AMSa3E

      \let\leq=\leqslant 
      \let\geq=\geqslant \let\ge=\geqslant
    \fi
  \fi
\fi 

\ifnfsstwo
  \DeclareMathAlphabet{\mathbfit}{OT1}{cmr}{bx}{it}
  \SetMathAlphabet\mathbfit{bold}{OT1}{cmr}{bx}{it}
  \DeclareMathAlphabet{\mathbfss}{OT1}{cmss}{bx}{n}
  \SetMathAlphabet\mathbfss{bold}{OT1}{cmss}{bx}{n}
  \ifAMStwofonts
    \ifCUPmtlplainloaded \else
      \DeclareSymbolFont{UPM}{U}{eur}{m}{n}
      \SetSymbolFont{UPM}{bold}{U}{eur}{b}{n}
      \DeclareSymbolFont{AMSa}{U}{msa}{m}{n}
      \DeclareMathSymbol{\upi}{0}{UPM}{"19}
      \DeclareMathSymbol{\umu}{0}{UPM}{"16}
      \DeclareMathSymbol{\upartial}{0}{UPM}{"40}
      \DeclareMathSymbol{\leqslant}{3}{AMSa}{"36}
      \DeclareMathSymbol{\geqslant}{3}{AMSa}{"3E}

      \let\leq=\leqslant 
      \let\geq=\geqslant \let\ge=\geqslant
    \fi
  \fi
\fi 

\ifCUPmtlplainloaded \else
  \ifAMStwofonts \else 
    \def\upi{\pi}
    \def\umu{\mu}
    \def\upartial{\partial}
  \fi
\fi

\title[The compact radio source PKS1549-79]
{The co-evolution of the obscured quasar PKS1549-79 and its host galaxy:
evidence for a high accretion rate and warm outflow}
\author[Holt et al.]
       {J. Holt$^{1}$, C. Tadhunter$^{1}$, 
       R. Morganti$^{2,3}$, M. Bellamy$^{1}$, R.M. Gonz\'alez Delgado$^{4}$, \newauthor
       A. Tzioumis$^{5}$, K.J. Inskip$^{1}$
	\\
$^{1}$Department of Physics and Astronomy, University of Sheffield,  Sheffield, S3 7RH, UK\\ 
$^{2}$ASTRON, PO Box 2, 7990 AA Dwingeloo, The Netherlands\\
$^{3}$Kapteyn Astronomical Institute, University of Groningen, P.O. Box 800, 9700 AV Groningen,
The Netherlands \\
$^{4}$Instituto de Astrofisica de Andalucia, Apdto. 3004, 18080 Granada, Spain \\
$^{5}$ATNF-CSIRO, Epping, Sydney, Australia\\}

\date{}
\def\ltsim{\ifmmode\stackrel{<}{_{\sim}}\else$\stackrel{<}{_{\sim}}$\fi}
\def\gtsim{\ifmmode\stackrel{>}{_{\sim}}\else$\stackrel{>}{_{\sim}}$\fi}
\begin{document}
\maketitle
\begin{abstract}{\large}
We use deep optical, infrared and radio observations to explore the symbiosis
between the nuclear activity and galaxy evolution in the southern
compact radio source PKS1549-79 ($z=0.1523$). The optical imaging observations reveal the
presence of tidal tail features which provide strong evidence that the host galaxy has undergone
a major merger in the recent past. The merger hypothesis is further supported
by the detection of a 
young stellar population, which, on the basis of spectral synthesis modelling of
our deep VLT optical spectra, was formed 50 -- 250~Myr ago and makes up a significant 
fraction of the total stellar mass (1 -- 30\%). Despite the core-jet structure of the radio source,
which is consistent with the idea that the jet is pointing close to our line of sight, 
our HI 21cm observations reveal significant HI absorption associated with both the
core and the jet. Moreover the luminous, 
quasar-like AGN ($M_V < -23.5$) is highly extinguished ($A_v > 4.9$) at optical
wavelengths  and shows many properties 
in common with narrow line Seyfert 1 galaxies (NLS1), including relatively narrow permitted lines 
($FWHM \sim$1940~km s$^{-1}$), highly blueshifted [OIII]$\lambda\lambda$5007,4959 lines ($\Delta V \sim 680$~km s$^{-1}$), and evidence that the putative supermassive black hole is accreting at 
a high Eddington ratio ($0.3 < L_{bol}/L_{edd} < 11$). 
The results suggest that accretion at high Eddington ratio does not 
prevent the formation of powerful relativistic jets. 

Together, 
the observations lend strong support to the  predictions of some recent numerical simulations
of galaxy mergers in which the black hole grows rapidly through merger-induced accretion
following the coalescence of the nuclei of two merging galaxies, and the
major growth phase is largely hidden at optical wavelengths by the natal gas and dust. 
Although the models also predict that AGN-driven outflows will eventually remove the gas from
the bulge of the host galaxy, the {\it visible} warm outflow in PKS1549-79 is not currently capable of
doing so. However, much of the outflow may be hidden by the material obscuring the quasar and/or
tied up in hotter or cooler phases of the interstellar medium.

By combining our estimates of the reddening of the quasar with the HI 
column derived from the 21cm radio observations 
we have also made the first direct estimate of the HI spin temperature in the vicinity 
of a luminous AGN: $T_{spin} > 3000$K.

\end{abstract}
\section{Introduction}
 
Powerful radio galaxies have long been considered as useful signposts to massive 
structures in the distant universe. However, there is now increasing speculation that AGN 
and associated jet activity are not merely signposts but, in fact, intimately linked to the 
galaxy evolution process. The main evidence for such links is twofold. On the one hand 
the shape of the redshift evolution in the number density of quasars and radio galaxies 
shows marked similarities to that of the star formation history of the universe
(Dunlop \& Peacock 1990, Madau et al. 1996), while on 
the other close correlations have been found between the masses of the supermassive 
black holes that produce the prodigious nuclear activity and the properties of the bulges 
of the host galaxies (e.g. Tremaine et al. 2002).

The correlations between the redshift evolution of the radio source 
population and the star formation history of the field galaxy population can be explained 
in terms of hierarchical galaxy evolution models (e.g. Kauffmann \& Haenelt 2000). 
It is well known from studies of nearby merging galaxies that stars are 
formed as part of the merger process. Numerical simulations of mergers of gas rich 
galaxies also show that the tidal torques associated with the mergers concentrate gas in 
the circumnuclear regions of the galaxies (Barnes \& Hearnquist 1996, 
Mihos \& Hearnquist 1996);  some of this gas may be available to fuel 
circumnuclear starburst, quasar and jet activity. 
Therefore it is likely that AGN activity and near-nuclear starbursts in massive galaxies are a 
direct consequence of hierarchical galaxy evolution; AGN 
may be a symptom of the overall galaxy evolution process.

In this context, radio galaxies are key objects for understanding the link between star 
formation and AGN activity in individual galaxies. Not only is there clear morphological 
evidence that a significant subset of powerful radio galaxies --- $\sim$50\% of those with strong 
emission lines --- have undergone recent mergers (Heckman et al. 1996, Smith \& Heckman 1989), 
but many of the radio galaxies with 
morphological evidence for mergers also show signs of recent star formation 
(Tadhunter et al. 1996, Aretxaga et al. 2001, Tadhunter et al. 2002, 2005, Wills et al. 2002, 2004,
Johnston et al. 2005). 
Despite these results, considerable uncertainties remain about the relationship between 
star formation and AGN/jet activity in radio galaxies. For example, it is not certain at 
what stage of the merger the AGN activity is triggered, nor whether the star formation is 
exactly coeval with the AGN activity. Indeed, some recent studies of the young stellar 
populations in radio galaxies suggest that the activity may be triggered relatively late in 
the merger sequence, a significant time after the major episode of merger-induced star 
formation (Tadhunter et al. 2005, Emonts et al. 2005, Johnston et al. 2005). However, the 
sample sizes are 
currently small and further detailed studies of 
the young stellar populations in powerful radio galaxies are required to put our 
understanding of the link between star formation and AGN activity on a firmer footing.

While the links between star formation and AGN/jet activity provide evidence that AGN 
are a bi-product of the overall hierarchical galaxy evolution process, it has been 
suggested that the correlations between the supermassive black holes masses and the 
properties of the bulges of the host galaxies are a consequence of the feedback effects of 
AGN-induced outflows (Silk \& Rees 1998, Fabian 1999, di Matteo et al. 2005).  
To date, most of the evidence for such outflows has been 
provided by observations of narrow and broad absorption lines observed in the spectra of 
AGN at UV and X-ray wavelengths (see Crenshaw, Kraemer \& George 2003 for a review). 
The outflows associated with 
these absorption features clearly have an impact on the near-nuclear environments 
on a scale $<$10pc,  but the impact they have on the 1 -- 10 kpc scale 
of galaxy bulges is less clear. Although outflows have also been detected in both ionized and 
neutral gas on a kpc-scale in powerful radio galaxies, uncertainties remain about the 
general importance of the outflows and their driving mechanism (quasars, jets or 
starbursts). 

The subject of this paper --- PKS1549-79 --- is a key object for studies of the links 
between galaxy evolution and AGN activity, because it shows clear evidence of both the 
merging process that may trigger the activity, and the feedback effect of the AGN activity 
on the surrounding ISM.

PKS1549-79 is a compact flat spectrum radio source. Previously unpublished VLBI radio 
images of the source show that it has the compact core-jet structure expected for such 
sources, consistent with the idea that its jet is pointing close to the line of sight. 
Remarkably for such a source, low-resolution optical spectra fail to reveal the 
broad permitted lines and strong non-stellar continuum characteristic of quasars and BL 
Lac objects (Tadhunter et al. 2001). Therefore this object does not readily fit in with 
the simplest versions of the 
unified schemes which propose that lines of sight close to the radio axis have relatively 
unobscured views of the AGN (Barthel 1989).

Despite the absence of strong quasar continuum and broad emission line features, 
optical spectra of the 
source do reveal  high ionization forbidden emission lines which suggest that the 
source contains a powerful quasar-like AGN, even if this AGN is not directly visible at 
optical wavelengths.  However, the high ionization emission lines are unusually broad, 
and blueshifted by $\sim$650 km/s relative to the low ionization lines. In addition, this 
object has HI 21cm absorption detected against its radio core, at the same redshift 
as the low ionization emission line system. Further support for the idea 
that the quasar nucleus is obscured by the circum-nuclear material in this source is 
provided by K-band spectroscopic observations that detect a broad Pa$\alpha$ emission 
line and strong non-stellar continuum characteristic of quasars (Bellamy et al. 2003).

Aside from its unusual emission line properties, PKS1549-79 belongs to the subset of 
$\sim$30 -- 50\% of powerful radio galaxies for which young stellar populations
make a substantial contribution to the optical continuum emission (Tadhunter et al. 2002). 
Further evidence 
for star formation in this object 
is provided by its unusually strong far-IR emission, which leads to its classification as 
an ultra luminous infrared galaxy (ULIRG: see Sanders \& Mirabel 1996 for a definition) with 
$L_{ir} =1.6\times10^{12}$~L$_{\odot}$ . This 
evidence for recent star formation is consistent with the idea that the host 
galaxy of this source has been involved in a merger 
which has also triggered a 
major episode of star formation.

In this paper we present new deep spectroscopic and imaging observations of PKS1549-79 
at optical, infrared and radio wavelengths that allow us to investigate the triggering of its 
AGN activity, characterise its circumnuclear outflows, and deduce the nature of its AGN. 
In this way we can place this key object in the context of evolutionary scenarios for the 
origin and evolution of the activity. 

We assume cosmological parameters of $H_0 =75$~km s$^{-1}$ and $q_0 = 0.0$ throughout the paper. 
For these parameters 1 arcsec corresponds 
to 2.39~kpc at the redshift of the host galaxy of PKS1549-79.

\section{Observations and reductions}

\subsection{Optical spectroscopy}

Optical spectroscopic observations were taken in 2002 July
and 2003 September using the EMMI spectrograph in RILD mode on the
ESO 3.5-m New Technology 
Telescope (NTT), and the FORS2 spectrograph in LSS mode on
the ESO Very Large Telescope (VLT). Both sets of data are deeper and 
have a higher spectral resolution
than those reported in Tadhunter et al. (2001).
Details of the observations are given in Table 1. 

On the NTT, grisms \#5 and \#6 were used with
the MIT/LL mosaic to obtain spectra with wavelength ranges
3215-6136\AA~and 4977-7565\AA~in the blue and red respectively.
These configurations  resulted in spectral resolutions of 5.0 -- 6.7\AA\, in the blue
and 4.2 -- 5.6\AA\, in the red, depending on slit width.
Red and blue
exposures were interleaved in an attempt to ensure 
uniformity in conditions between the exposures. To reduce the effects
of differential refraction, all exposures were taken with the slit
aligned along the parallactic angle. Spectra were taken along two slit
positions -- PA 25 and PA -5. 
 
The aim of the NTT spectra was to investigate the kinematics and
physical conditions of the emission line gas through modelling of the
bright emission lines. However, higher signal-to-noise is required to
investigate the stellar populations in PKS 1549-79. Hence,
we have also obtained optical spectra with the FORS2 spectrograph on the
VLT on 2003 September 25. Grisms 600B and 600RI (holographic) were
used with the MIT CCD to obtain spectra with wavelength ranges
3100-6000\AA~and 5000-8200\AA respectively.  The  
VLT spectra were 
taken with a 1.3 arcsec slit to ensure the entire nuclear region was
sampled, and the slit was aligned along the parallactic angle (PA 75). The
resulting spectral resolutions were 
6.5\AA\, and 7.4\AA\, for the blue and red spectra respectively.
As with the NTT observations, the red and blue
exposures were interleaved.

For both the NTT and the VLT observations the 
data reduction followed the standard steps of bias subtraction, flat fielding, 
wavelength calibration, atmospheric extinction correction, flux calibration, and
registration of blue and red spectral images. Based on measurements of night sky lines,
the uncertainty in the wavelength calibration is estimated to be smaller than
0.5\AA\, for all datasets; the relative flux calibration uncertainty, based on 
the comparison of observations of several spectroscopic standard
stars, is better than $\pm$5\%.
Prior to the analysis of the data, the spectra were corrected for a Galactic
extinction assuming $E(B-V)=0.679$ (Schlegel, Finkbeiner \& Davis 1998) and the Seaton
(1979) extinction law.

\begin{table}
\begin{tabular}{lclcc}\\  \hline
Date &Exposure  &PA &Slit width &Seeing FWHM  \\
&(s) & &(arcsec) &(arcsec) \\ \hline
\multicolumn{5}{l}{\bf NTT/EMMI (Optical Spectroscopy)} \\
12/07/02  &3*1200(R)  &-5 &1.0 &0.78 \\
  &3*1200(B)  & -5 &1.0 &0.78 \\
13/07/02 &2*1200(R)  & 25 &1.5 &0.78 \\
 &2*1200(B)  &25 &1.5 &0.78 \\
\\
\multicolumn{5}{l}{\bf VLT/FORS1 (Gunn~r Imaging)} \\
25/07/03 &4*300  &- &- &0.67 \\
\\
\multicolumn{5}{l}{\bf VLT/FORS2 (Optical Spectroscopy)} \\
25/09/03  &3*600(R)    &75 &1.3 &1.2-1.8 \\
25/09/03 &3*1200(B)    &75 &1.3 &1.2-1.8 \\ 
\\
\multicolumn{5}{l}{\bf AAT/IRIS2 (K-band Spectroscopy)} \\
4/09/03 &24*120   &0 &1.0 &1.0 \\
\\
\multicolumn{5}{l}{\bf NTT/SOFI (H and K-band Spectroscopy)} \\
24/08/04 &12*120  &0 &1.0 &1.2 \\
\end{tabular}
\caption{Log of
optical and infrared observations for PKS1549-79. 
In the case of the optical spectroscopic
observations the seeing was estimated from the DIMM monitor estimates at the time of the observations,
whereas for the imaging observations the seeing was measured directly from star images in the field of 
the
source.}
\end{table}

\subsection{Optical imaging}

Deep images of PKS1549-79 were obtained using FORS1 on the VLT with a Gunn~r filter in
good seeing conditions ($FWHM=0.67$ arcseonds).
The advantage of this filter is that, covering the rest wavelength range $\sim$5200 -- 6100\AA, 
it avoids all strong emission lines in the source and is therefore dominated
by continuum light. During the
observations the telescope was jittered to four separate positions on the sky to help remove cosmetic
defects and average out sensitivity variations.

The imaging data were reduced using a combination of the STARLINK FIGARO package and IRAF, by first
subtracting the bias, dividing each separate frame by a jittered sky flat field, then combining
the images with IMCOMBINE in IRAF using the coordinate information in the image headers.

\subsection{Near-IR spectroscopy}

In an attempt to improve on the near-IR spectroscopic data of
Bellamy et al. (2003), which were affected by a faulty grism in IRIS2,
new near-IR spectra of PKS 1549-79 were taken with the Anglo Australian Telescope (AAT),
and the NTT. Details of the observations are given in Table 1.

The first data set was obtained on 2003 September 4 using the
IRIS2 instrument on the AAT. The Sapphire
240 grism was used to cover the K band. In total, twenty four
exposures of 120s were obtained for PKS 1549-79, with the
galaxy  `nodded' between two positions on the slit in an ABBA pattern.
Four exposures of 66s were taken of the A0
star HIP 77941 at a similar airmass for calibration purposes.

The second set was obtained on 2004 August 24 with the SOFI
instrument on the NTT. The low resolution
red grism was used with the GRF order-sorting filter 
to cover the H and K bands simultaneously. Cloud was present during the
night and the observations were not photometric. 12 exposures of 120s were
obtained; the galaxy was again nodded along the slit but this time
with a random jitter every 2 exposures. Eight exposures of 30s were
taken of the A0 star HIP 79933 at a similar airmass for calibration
purposes.

The IRIS2  
exposures were co-added using median filtering to
remove cosmic rays; the median `B' image was then subtracted from the
median `A' image to remove the night sky lines. The standard star
frames were combined in a similar way. The resulting galaxy and star
frames were then flat-fielded using a dark-subtracted flat-field taken
with the same instrumental setup. The night sky lines from the data frames
themselves were used to determine a two-dimensional wavelength
calibration, which was then applied to the final co-added frames of the
galaxy and the standard star. The resulting spectra cover the range
20320 -- 24900\AA. The spectral resolution,
measured from the night sky lines, is 10.9$\pm$0.4~\AA(FWHM). Fits to
the night sky lines show that the wavelength calibration is accurate to
$\pm$0.16~\AA. The two-dimensional 
spectra were straightened using the APSUM package in IRAF, and
one-dimensional spectra were extracted with FIGARO.

Night sky lines were also used to derive a wavelength calibration for
the NTT data. The 
resulting two-dimensional transformation was then
applied to all the individual data frames and then suitable groups of
exposures (because of the random jittering) were combined with FIGARO
and a series of one-dimensional spectra were extracted and then
combined together. There was no appreciable tilting of the slit
image and so the SOFI spectra did not need to be straightened before
extraction. The SOFI spectrum covers the wavelength range 15057 -- 25392\AA.
Atmospheric absorption dominates between $\sim$18000 and 19500\AA\, 
and the thermal background dominates at wavelengths greater
than $\sim$24000\AA. The spectral resolution, measured from the
arc lines, is 35.0$\pm$0.4~\AA(FWHM). Fits to the night sky
lines show that the wavelength calibration is accurate to $\pm$ 0.77
\AA.

The standard star exposures were used to remove telluric features
and also to flux calibrate the IRIS2 spectra. The magnitude-to-flux
conversion was performed with reference to Bessell, Castelli \& Plez
(1998), while telluric correction (and flux calibration for the IRIS2 data) was
performed by assuming the intrinsic SED of the A0 stars to be that of
a black body at $T = 9480$K. For hot stars this is a good approximation,
particularly in the near-IR end of the black body spectrum. A number
of Brackett lines are visible in the spectra of A0 stars, particularly
in the H band, but these were masked out before the telluric
correction was performed.

The SOFI data suffered unquantifiable slit-losses
due to the
telescope nodding axis not being aligned precisely with the slit. For
this reason no attempt was made to make an absolute flux calibration of 
the SOFI spectrum. Instead the spectrum was scaled to match the
IRIS2 spectrum in the K-band and the spectrum of HIP 79933 was used
only to remove telluric features and obtain the correct spectral shape.

\subsection{Radio observations}

Continuum VLBI observations of PKS1549-79 were made at both 2.3 and 8.4GHz with 
the SHEVE network (Preston et al. 1993, Jauncey et al. 1994) between November 1988 
and March 1992, using the MK2 recording system (Clark 1973). Full details of the 
antennae used for SHEVE observations are given in Tzioumis et al. (2002). The tapes 
recorded at each station were processed at the Caltech-JPL Block2 correlator in Pasadena, 
and the NRAO AIPS package was used for global fringe fitting. The amplitude 
calibration and editing of the visibilities were subsequently performed using the Caltech 
VLBI package. Self-calibration and imaging of the data were performed using DIFMAP 
(Shepherd et al. 1994) and the AIPS packages. At 2.3GHz the restoring beams were
 7.3$\times$2.7~mas (PA4.7) and 8.9$\times$7.2~mas (PA9.8) with and without the Harteebeesthoek antenna respectively, whereas at 8.4~GHz the restoring beam was 6.7$\times$5.2~mas (PA73) 
without the Hartebeesthoek antenna.  The r.m.s. noise in the final images
is  2.7 mJy beam$^{-1}$   at 2.3~GHz, and 3.8 mJy beam$^{-1}$ at 8.4~GHz.

In order to investigate the spatial distribution of the HI 21cm absorption previously 
detected by Morganti et al. (2001), HI VLBI observations were carried out using the Long Baseline Array (LBA) 
at the redshifted HI frequency of 1645~MHz. At this relatively low frequency we could 
use only three telescopes: Parkes (64m), ATCA (Tied Array, 5$\times$22m) and Mopra (22m). 
We used a 16MHz bandwidth in each circular polarization and 1024 channels. The final data 
cube (after Hanning smoothing) has a spectral resolution of $\sim 9$ km s$^{-1}$, and a noise per
spectral channel of 5.3 mJy beam$^{-1}$, while the continuum map derived from these data has a noise of
16 mJy beam$^{-1}$.

The editing part of the calibration of the 21cm line data was done in AIPS and then the 
data were transferred to MIRIAD (Sault, Teuben \& Wright 1995) for the bandpass 
calibration. The calibration of the bandpass was done using additional calibrations of the 
strong calibrators PKS1718-649 and PKS1934-638.

Both the line and the continuum image were made using uniform weighting and the 
restoring beam is 47$\times$32mas (PA-77.3). As expected, the resolution is lower than 
that obtained in the previous VLBI continuum observations described above. The total 
cleaned flux in the continuum image is 4.75 Jy --- close to that obtained from the 
previous low resolution ATCA observations.

\section{Results}

\subsection{Optical imaging}

\begin{figure*}
\epsfig{file=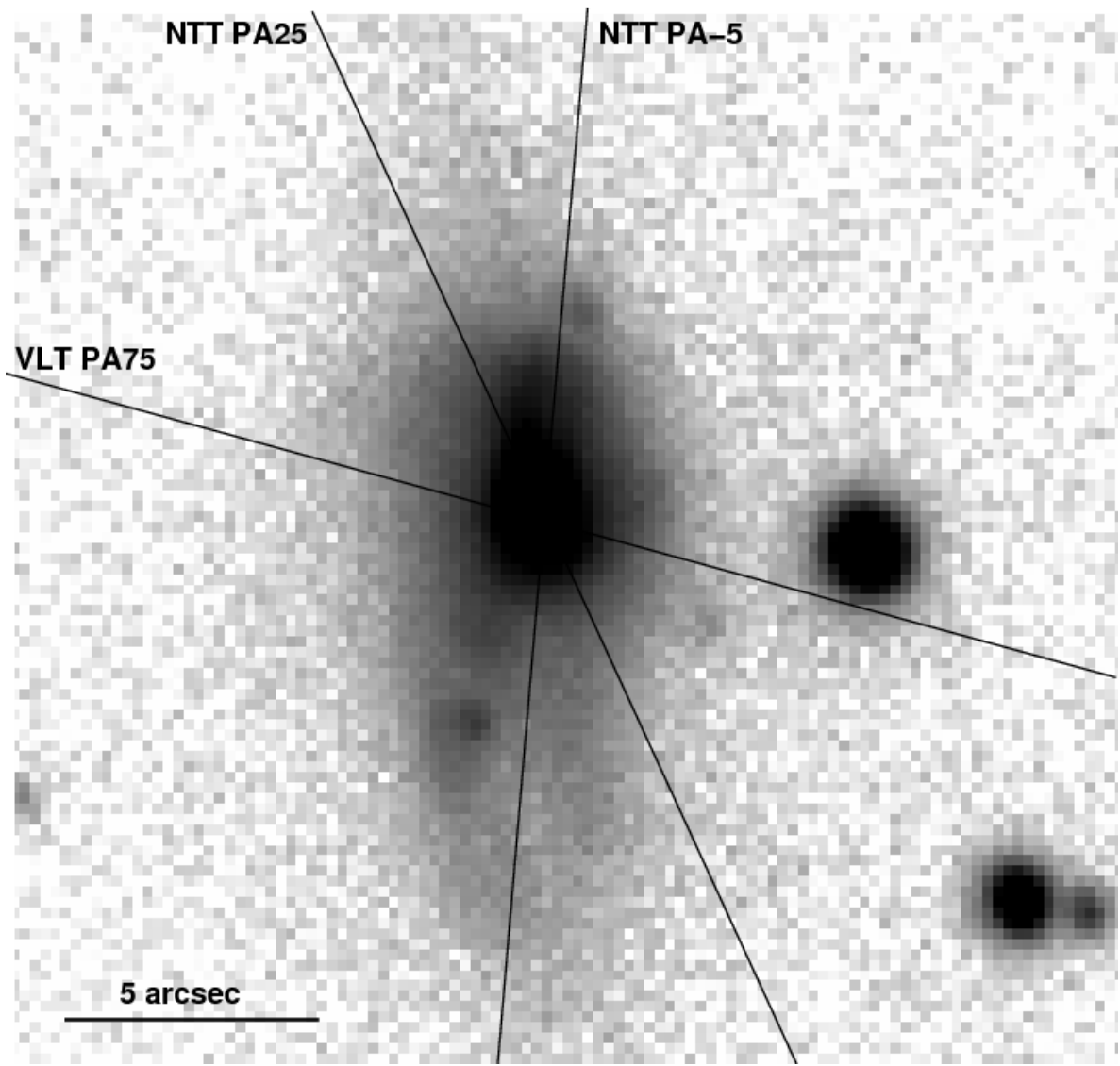,width=13cm,angle=0}
\caption{Deep VLT image of PKS1549-79 taken with a Gunn~r 
filter that avoids strong emission lines from the galaxy. The slit positions 
used for the long-slit spectroscopic observations are shown for reference. North
is to the top and east to the left.}
\end{figure*}

Our deep VLT image of PKS1549-79 is shown in Figure 1. This reveals a 
disturbed optical continuum morphology, with a high surface 
brightness elongation extending 4 arcseconds (10~kpc) to the north of the nucleus, a 
pair of tidal tails (or loop feature observed close to edge-on)
extending a total distance of 10 arcseonds (24~kpc) to the south, and a fainter jet-like
feature that extends 12 arcseconds (29~kpc) to the NNE of the nucleus. The extended structures
contain several knots of emission,  particularly
to the SSE and NNW. Overall, the structures suggest that PKS1549-79 has undergone
a recent merger, which induced star formation in the extended halo of the galaxy.

\subsection{Optical spectroscopy}

Three slit positions used for the optical spectroscopy observations are shown superimposed on the
Gunn~r image in Figure 1.

\subsubsection{Optical continuum modelling}

Accurate modelling of the continuum is required not only to deduce the properties of any young stellar 
populations present, but also as a prerequisite for accurate measurement of the emission lines, 
following subtraction of the best-fit continuum model. In this section we concentrate on the VLT 
spectrum extracted for the nuclear region because this has higher S/N in the continuum than either of 
the NTT spectra. 

Prior to continuum modelling, the level of the nebular
continuum was deduced following the procedure outlined
in Dickson et al. (1995) and Holt et al. (2003), based on double
Gaussian fits to the H$\beta$ emission line feature. This nebular continuum 
was then subtracted from the data. The major uncertainty in this
procedure is the reddening in the nebular continuum. Unfortunately, estimates of the reddening using
the Balmer lines are highly uncertain in this object (see section 3.2.3). To consider the effects of the 
reddening of the nebular continuum on the results of the continuum modelling, we carried out the 
continuum modelling twice with two extreme sets of assumptions about nebular continuum. First we 
assumed zero reddening for the nebular continuum. Second we assumed the maximum reddening allowed by 
the Balmer line ratios. 

The technique used to model the continuum is described in detail in Tadhunter et al. (2002, 2005). To 
summarise, the continuum flux is measured in several wavelength bins (in this case 39) chosen to 
avoid emission lines, cosmetic defects and poorly-corrected atmospheric absorption features. A minimum 
$\chi^2$ technique is then used to compare the measured fluxes with the results of spectral synthesis 
models. Models comprising various
combinations of an old stellar population (12.5~Gyr), young stellar populations (0.01 -- 2.5~Gyr),
reddened young stellar populations (0.01 -- 2.5~Gyr, $0.0 < E(B-V) < 2.0$), and a power-law are 
compared with the data. For the spectral synthesis results we used the most recent set of
high resolution models from Bruzual and Charlot (2003) for an instantaneous burst, a Salpeter IMF and 
solar metallicity. For the reddened young stellar populations (YSP) we used the Seaton (1979) 
extinction law to redden the YSP spectra from the spectral synthesis models. We also assumed a uniform 
uncertainty of $\pm$5\% for all the measured fluxes, consistent with the estimated uncertainty of the 
relative flux calibration. Based on our experience of this modelling technique, models with 
$\chi^2_{red} < 1$ can be deemed to provide a good fit to the data (see Tadhunter et al. 2005 for 
discussion).

In our first attempt to model the nebular-subtracted optical continuum, we considered the full 
wavelength range covered by the spectra. It became clear that no combination of old and young 
stellar populations (reddened or otherwise) could fit the relatively steep rise in the spectrum to the 
red at wavelengths longer than $\sim$6500\AA, and that a red power-law is required to provide an adequate 
fit in that spectral range. Note that the steep rise to the red is not due to any problem with the 
reduction and calibration of the data, since it is also present in our NTT spectra. Figure 2 shows 
the best fitting model that comprises a 12.5~Gyr old stellar population, a 0.05 Gyr young stellar 
population reddened by $E(B-V)=0.4$, and a power-law. The level of the power-law component is consistent
with the extrapolation of the reddened quasar continuum detected at near-IR wavelengths (see section
3.3). Indeed the existence of the red power-law provides independent evidence for the partially extinguished quasar 
first detected by Bellamy et al. (2003). 

\begin{figure*}
\epsfig{file=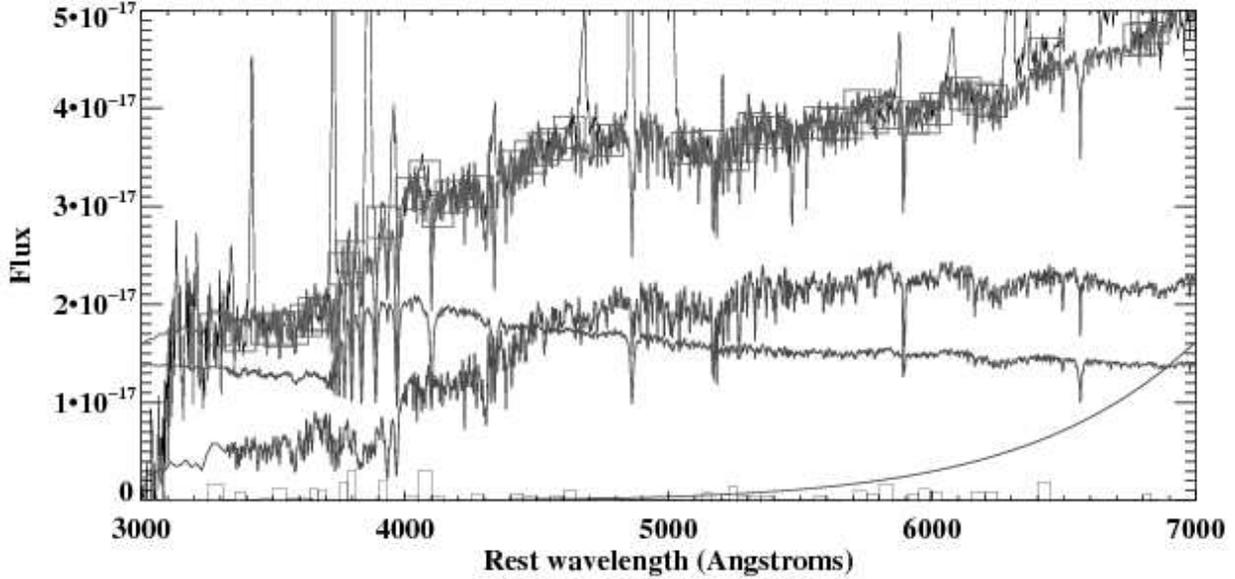,width=16cm,angle=0}
\caption{Spectral synthesis model fit to the full wavelength range of the
optical spectrum of PKS1549-79. The model comprises a 0.05Gyr YSP reddened by $E(B-V)=0.4$, a 12.5Gyr old stellar
population, and a power-law. Note that the power-law component is required to fit the steep rise in the continuum at the
far red end of the spectrum. The differences between the model and the data are presented as a histogram at the bottom
of the plot.}
\end{figure*}

\begin{figure*}
\begin{tabular}{cc}
\epsfig{file=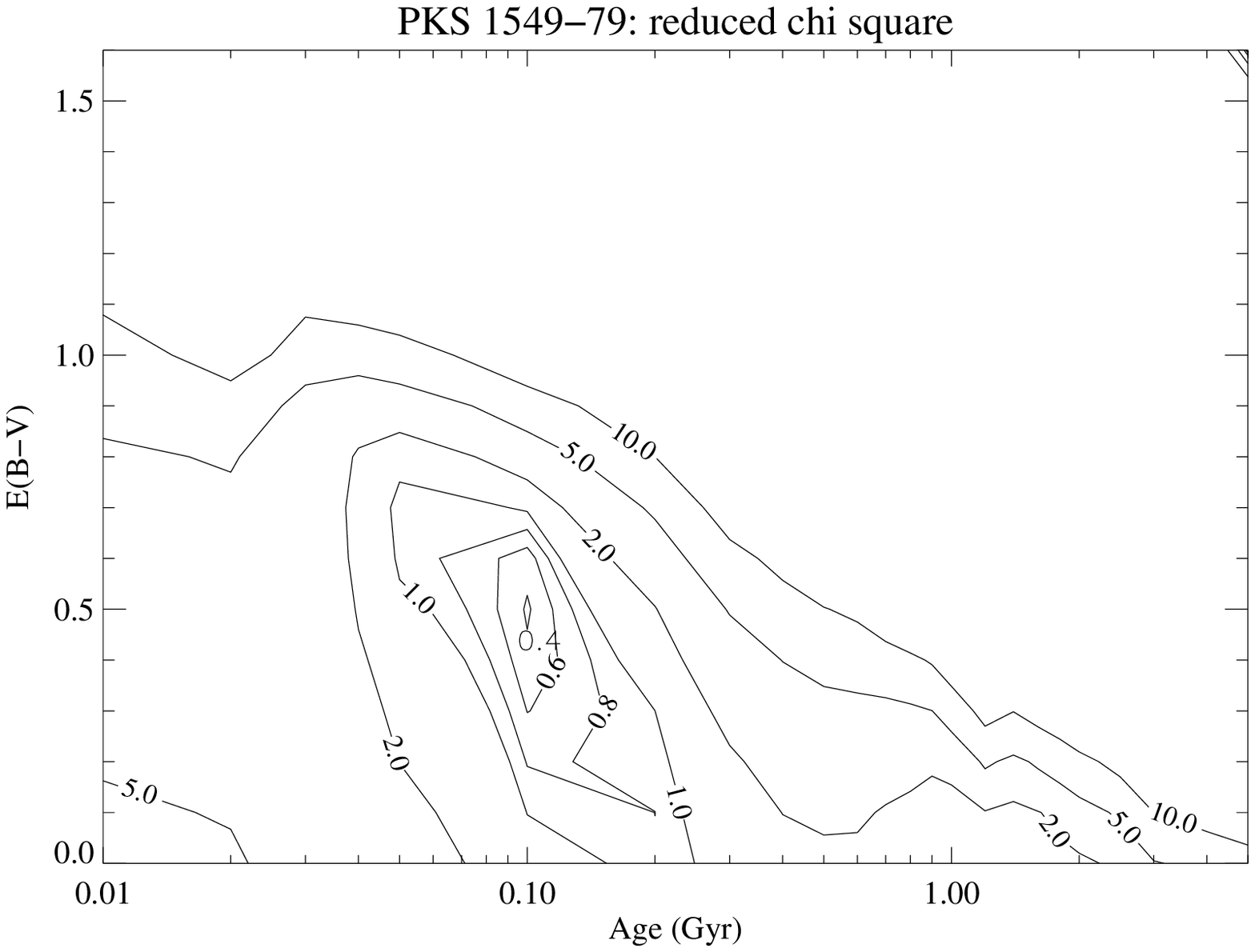,width=8cm,angle=0} &
\epsfig{file=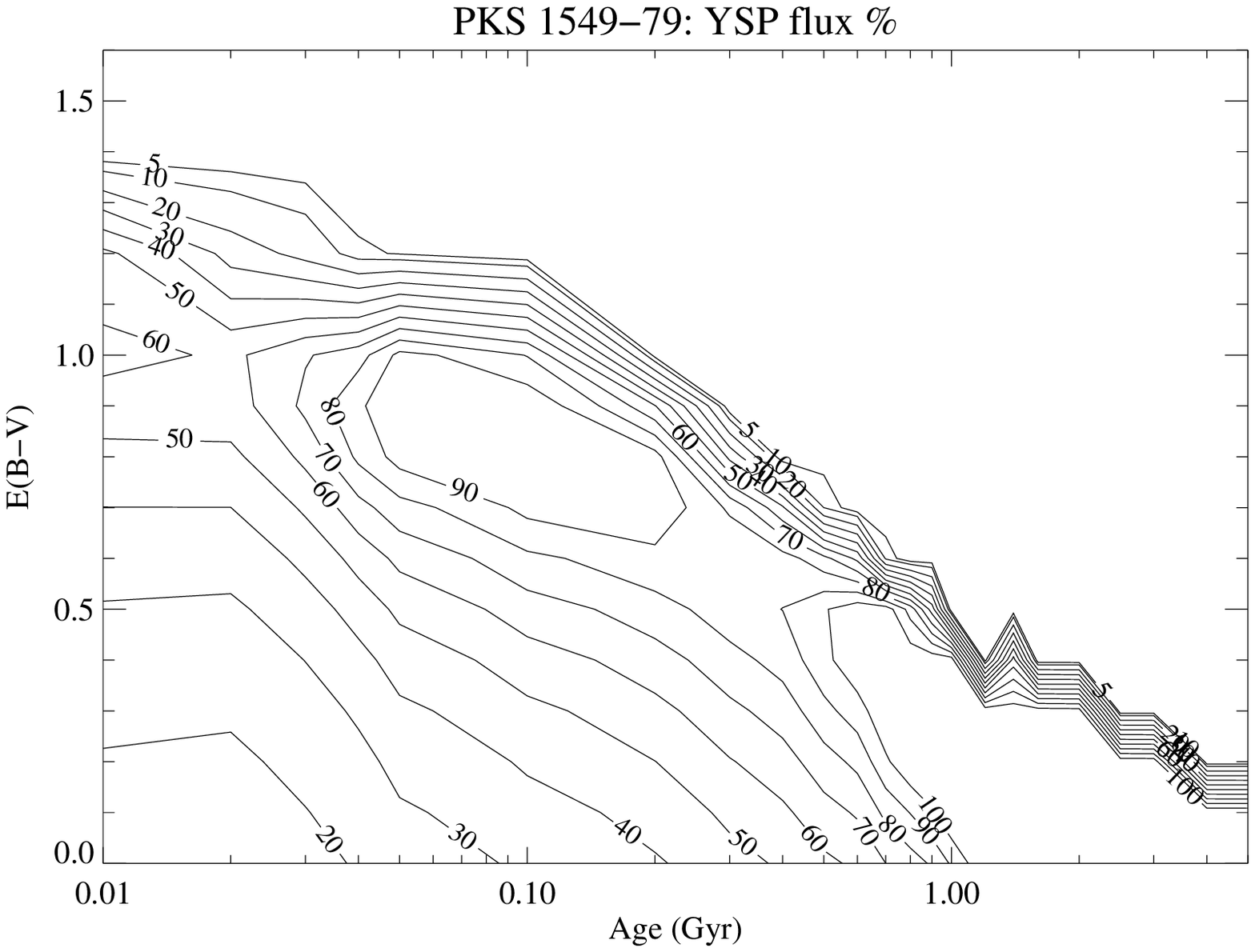,width=8cm,angle=0} \\
\end{tabular}
\caption{The results of minimum $\chi^2$ fitting to the optical continuum of PKS1549-79 assuming
the zero reddening for the nebular continuum, restricting the wavelength range of the fits to
$\lambda_{rest} < 6000$\AA. The plot on the left gives the reduced $\chi^2$ of the model fits for different combinations
of YSP age and reddening, while the plot on the right gives the percentage contribution of the YSP component in
the normalising bin used for the modelling ($4720$ -- $4820$\AA). Reduced $\chi^2$ values less than unity are deemed acceptable.}
\end{figure*}

\begin{figure*}
\epsfig{file=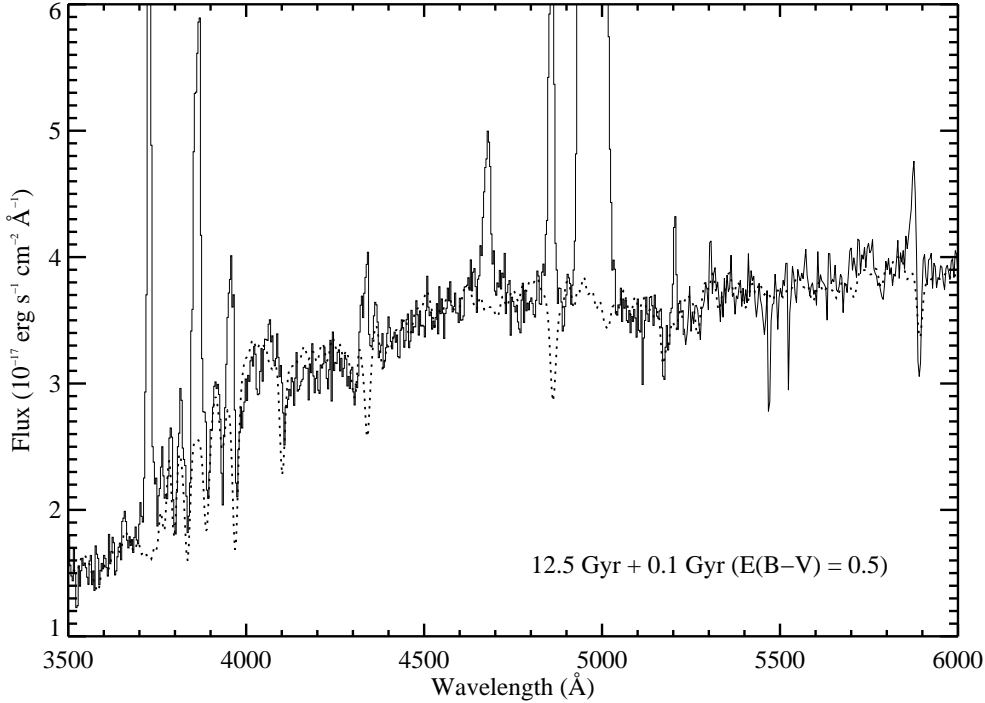,width=13cm,angle=0}
\caption{Example fit to the optical continuum for the model including a YSP with age
0.1~Gyr and reddening $E(B-V) = 0.5$ (zero reddening for the nebular continuum). The CaII~K, G-band and MgI absorption
features are well fitted by this model, but the higher order Balmer lines are likely to suffer from residual emission 
line contamination.}
\end{figure*}

\begin{figure*}
\begin{tabular}{cc}
\epsfig{file=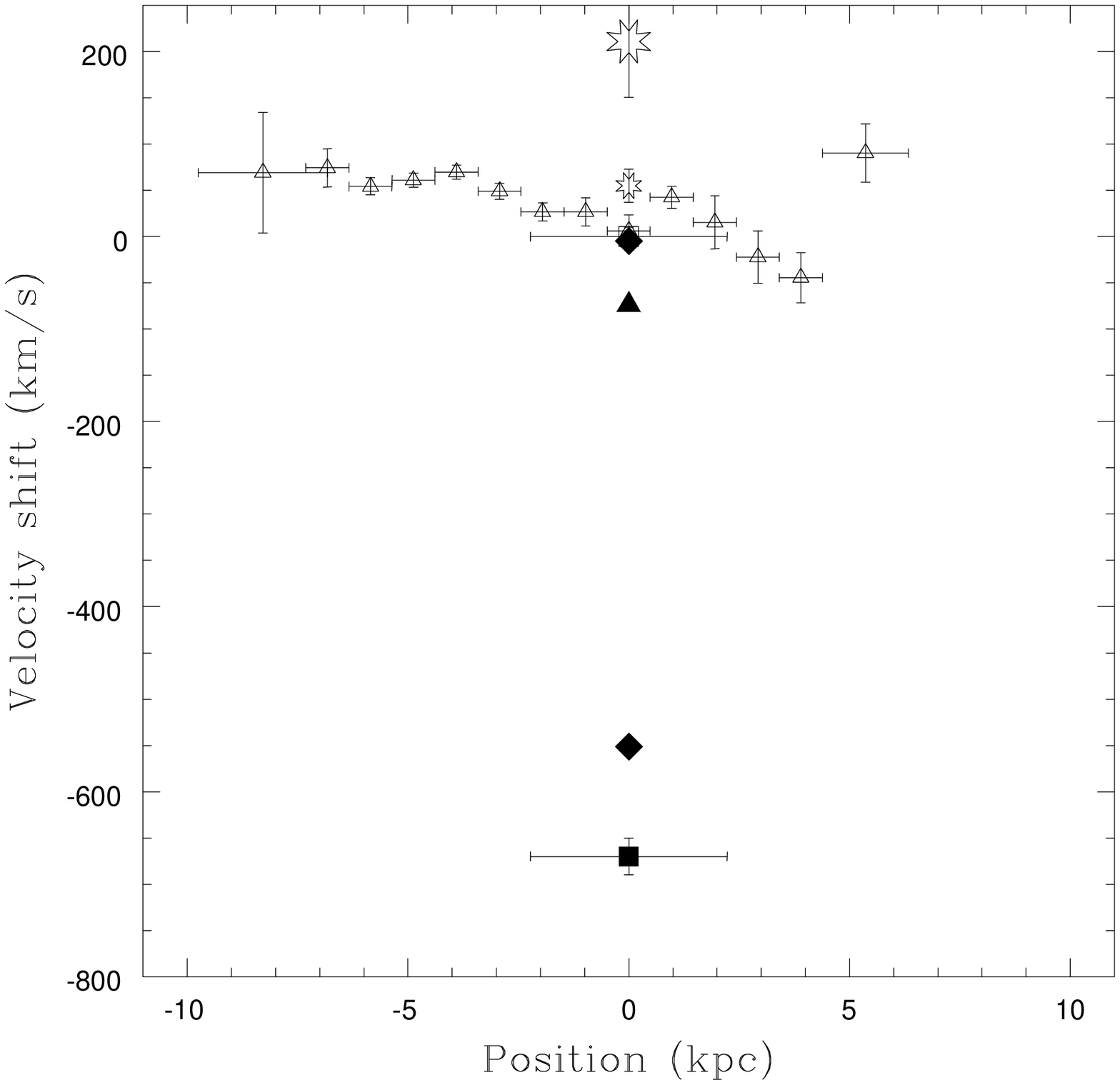,width=8cm,angle=0.} &
\epsfig{file=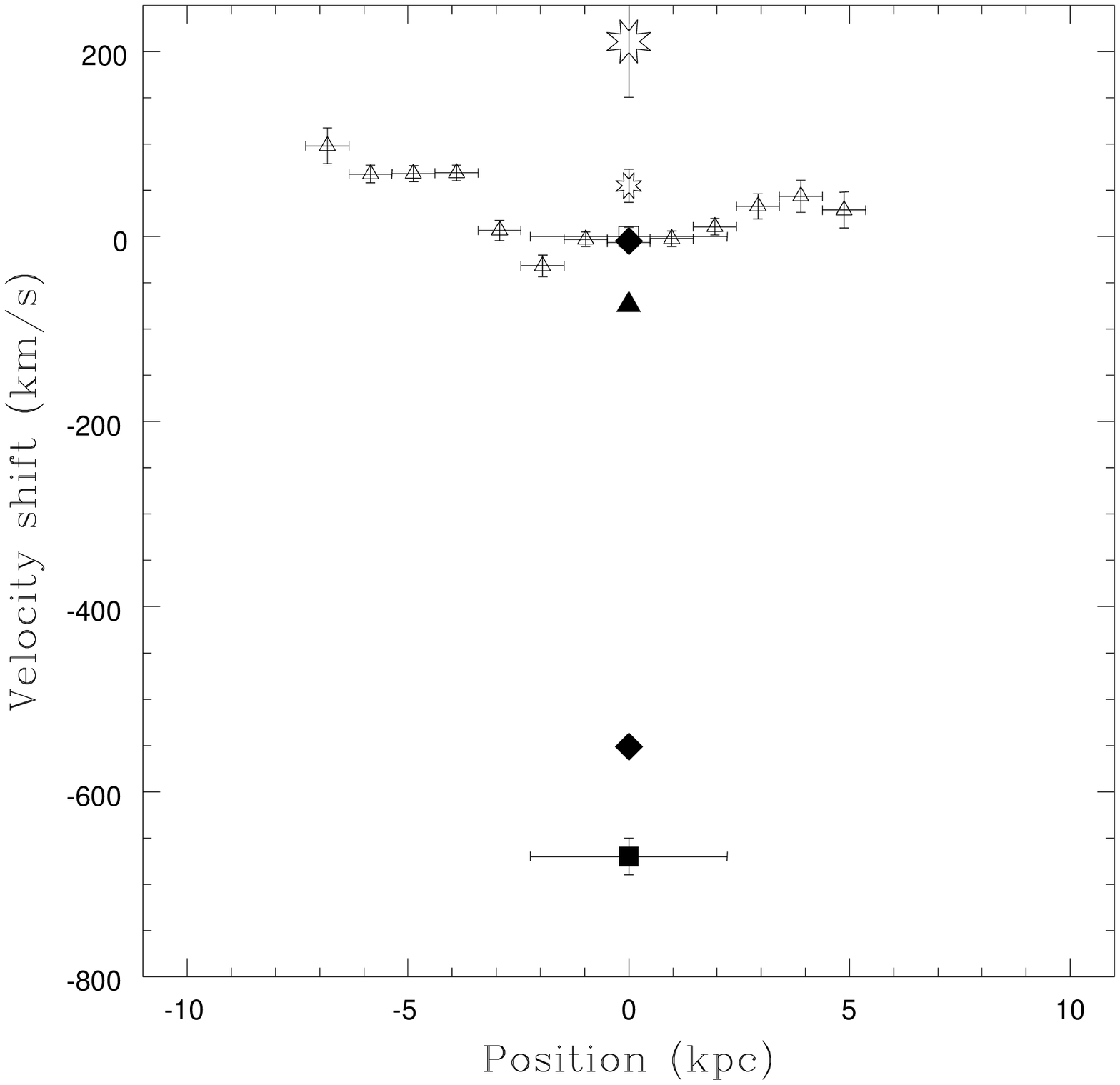,width=8cm,angle=0.}\\
\epsfig{file=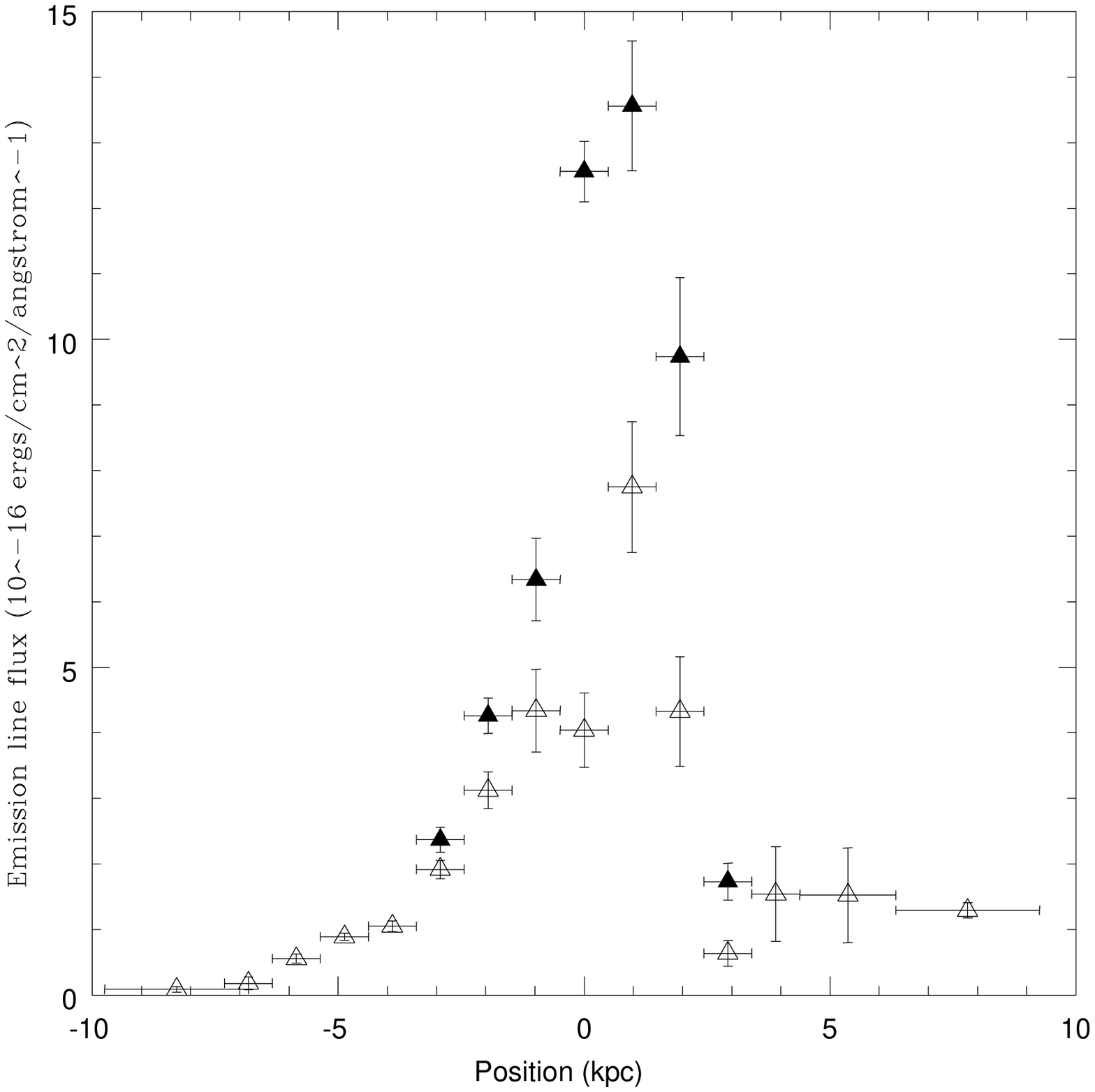,width=8cm,angle=0.} &
\epsfig{file=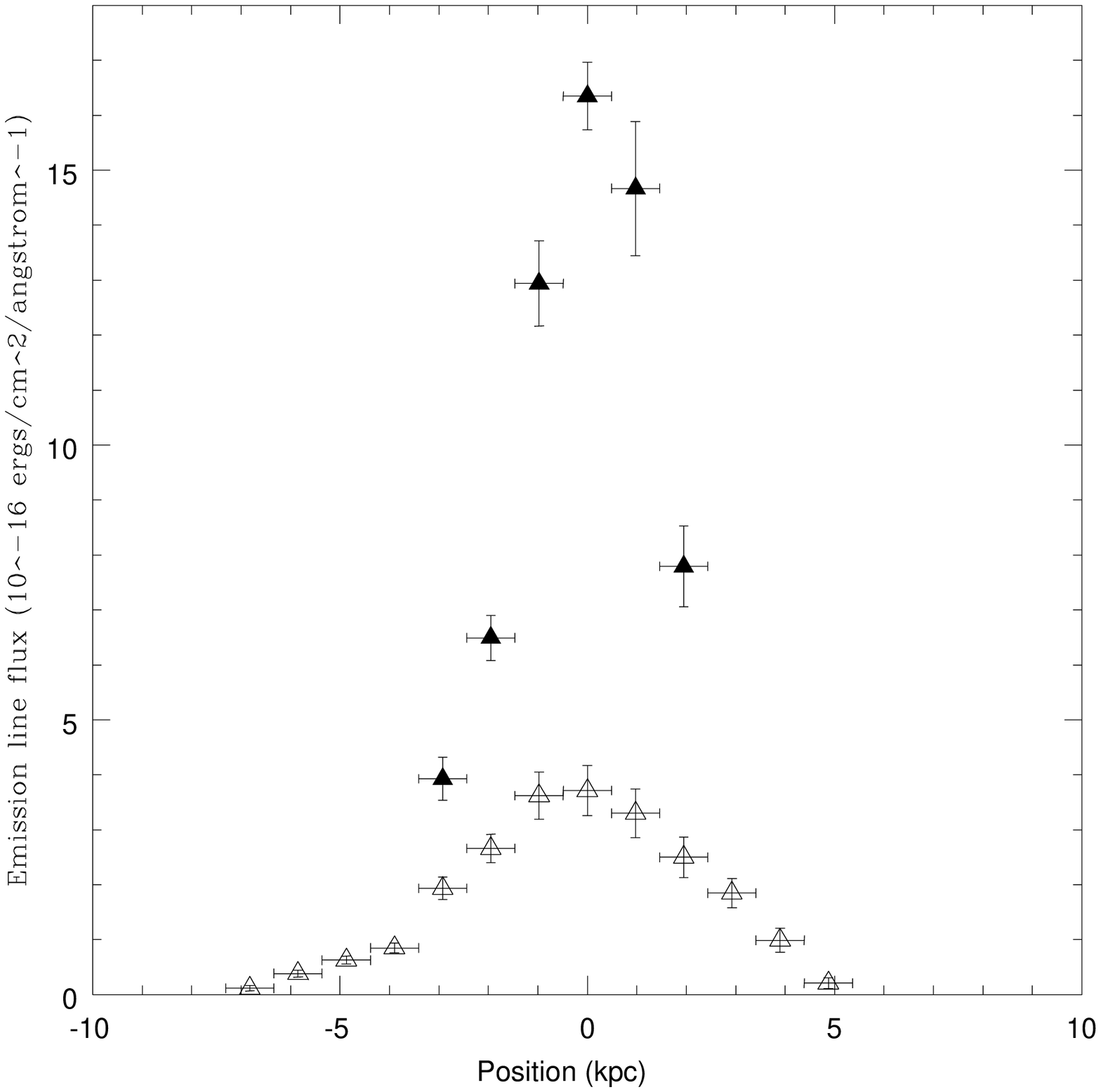,width=8cm,angle=0.}\\
\end{tabular}
\caption
{Radial velocity profiles (top) and the spatial variations of emission line
  flux (bottom) for the NTT observations of PKS 1549-79. The left hand images
  are for PA-5 and the right hand images are for PA25. 
Small open
  and filled triangles represent the narrow and intermediate components of
  H$\alpha$ respectively. 
Overplotted is the radial velocity of the deep HI 21cm
  absorption (large filled triangle at -30~km s$^{-1}$
  from Morganti et al. 2001), the shifts 
  of the [O II] and [O III] lines from Tadhunter et al. (2001)
  (large filled diamonds at -5~km s$^{-1}$ and -550~km s$^{-1}$ respectively), the 2
  components of the [O III] model in the 
  nucleus (large open and filled squares) and the broad components of Pa$\alpha$ listed
  in Table 2 (open
  8-pointed stars). For PA-5, positive
  positions are to the N and negative positions are to the S. For PA25, 
positive positions are to the SSW and negative positions are to
  the NNE.} 
\end{figure*}
In order to avoid the added uncertainly that the inclusion of a power-law 
introduces into the modelling of the continuum, and to deduce an accurate age for the YSP, 
we then excluded wavelengths $\lambda > 6000$\AA, from the model fits.  In the case of maximum reddening
for the subtracted nebular continuum, it is clear that the data are well fitted by models that include a combination 
of a 12.5~Gyr old stellar population and a YSP with age in 
the range 0.04 -- 0.1~Gyr and reddening in the range $0.0 < E(B-V) < 0.75$, with a best fit obtained 
for a YSP age of 0.05~Gyr and reddening of $E(B-V) = 0.6$. The results appear to be robust in the sense that, if we 
assume that the nebular continuum suffers no reddening, the results are similar, except that the range of acceptable 
models extends to older ages for the YSP (0.05 -- 0.25~Gyr) and the best fitting model has an age of 
0.1~Gyr and a reddening of E(B-V)$\sim$0.5. The results of the SED
modelling for the restricted wavelength range are shown in Figure 3 for the zero reddening in the 
nebular continuum.

The next step in the analysis was to check the SED fitting results by determining how well the models
fit the profiles of absorption lines that do not suffer from significant emission line contamination, 
namely the CaII~K, the G-band, and MgI features. Reassuringly, the models that fit the general SED 
well also provide the best fits to the absorption lines.  Figure 
4 shows an example of a fit to the optical spectrum for a model with a YSP of age 0.1~Gyr and 
reddening $E(B-V) = 0.5$, assuming zero reddening for the nebular continuum.

To summarise: fits to the optical continuum require a young stellar population (YSP) with age in the range 0.04 -- 0.25~Gyr
and reddening $0.0 < E(B-V) < 0.8$ (depending on age) in order to provide an adequate fit to the optical SED and stellar absorption 
features. The YSP has a mass $2\times10^8 < M_{ysp} < 4\times10^9$~M$_{\odot}$
and makes up 1 --- 30\% of the total stellar mass in the slit (depending on the exact age and reddening).
Note the age deduced for the YSP is significantly smaller than the 1~Gyr determined by Tadhunter et al. (2002) 
using lower resolution data with a narrower wavelength range. Much of this difference is likely to be due to the fact
that Tadhunter et al. (2002) did not correct their spectrum for Galactic extinction prior to the modelling.

\subsubsection{Emission line kinematics}

The long-slit spectra of PKS1549-79 provide useful information on the emission line
kinematics in both the nuclear and off-nuclear emission line regions. The spectra were measured by using the DIPSO package to fit Gaussian profiles to the lines. The velocity 
widths derived from the fits were
quadratically corrected for the instrumental profiles, and all line widths and radial
velocity shifts were corrected to the rest frame of PKS1549-79.  

The NTT spectra, which were obtained under good seeing conditions, provide the best information 
about the spatial extent and kinematics of the extended emission line gas. Figure 5 presents  
flux and radial velocity information derived from the fits to the emission line profiles. 
The H$\alpha$ emission line is detected across the 
full extent of the higher surface brightness continuum structures detcted in the Gunn~r continuum 
image shown in
Figure 1, extending 10 arcseconds to the south and 5 arcseconds to the north of the nucleus. The 
emission line
kinematics are relatively quiescent in the extended regions with small velocity amplitudes ($\Delta V 
< $150~km
s$^{-1}$) and linewidths (FWHM $<$  400 km s$^{-1}$) --- consistent with a gravitational origin for 
the gas 
motions (Tadhunter et al. 1989, Baum et al. 1990).

In contrast to the extended regions, the kinematics in the nuclear region provide evidence for
more extreme gas motions, as already noted by Tadhunter et al. (2001). In order to analyse the gas 
motions in the
spatially unresolved nuclear regions we have adopted the approach used by Holt et al. (2003) for the 
radio
galaxy PKS1345+12. This consists of fitting the minimum number of Gaussian components required model 
each line adequately. For
this analysis we have concentrated on the VLT data which have a higher S/N in the nuclear regions. In
the case of emission line doublets ([OIII]$\lambda\lambda$5007,4959, [OI]$\lambda\lambda$6300,6363,
[NII]$\lambda\lambda$6548,6584, [SII]$\lambda\lambda$6717,6731) the separations, and, with the 
exception of
the [SII] blend, the ratios of the lines were set by atomic physics, and it
was assumed that both components of the doublets have the same kinematic components.
We found
that for all the low ionization lines  [OII]$\lambda$3727, H$\beta$, [OI]$\lambda\lambda$6300,6363, 
[NII]$\lambda\lambda$6548,6584, 
and [SII]$\lambda\lambda$6717,6731, a double Gaussian fit comprising narrow (FWHM $<$ 400 km s$^{-1}$) 
and intermediate width blueshifted components (FWHM $>$ 1000 km s$^{-1}$) provided an adequate fit to the 
emission line 
profiles. 
For these low ionization lines the narrow component is strongest, and the widths and redshifts of the narrow 
components are 
consistent within the uncertainties for all the narrow lines; they are also consistent with the 
redshift of
the H$\alpha$ emission line detected in the extended regions (see Figure 5). The intermediate emission line component is 
relatively 
weak in the low ionization lines, and its width and redshift less well determined based on free double 
Gaussian fits.

Initial attempts to fit the [OIII]$\lambda\lambda$5007,4959 lines showed that it was possible to fit 
each of
line of the doublet adequately using a single Gaussian component that is both broad and blueshifted 
relative to the
narrow component detected in the low ionization lines (see also Tadhunter et al. 2001). However, given 
that 
results obtained for [OIII] in apertures slightly offset from the nucleus show  the
clear presence of a narrow (FWHM $<$ 400 km s$^{-1}$) emission line component that
is also detected in the low ionization lines, we have also 
fitted
each component of [OIII] doublet in the nucleus using a double Gaussian model, with one of the 
Gaussian components constrained to
have the velocity width and redshift of the narrow component detected in the low ionization lines. The 
resulting fit
--- comprising the narrow component and an intermediate, blueshifted component --- is shown in Figure 6. 
The
imperfections of this fit in the line cores are to likely reflect the fact that a single Gaussian is 
only an
approximation to the true profile of the intermediate, blueshifted component in the nucleus. Note that
such imperfections would not be apparent in fits to the low ionization lines, because of the weakness 
of the broad
components, and the dominance of the narrow components in such lines. 

Following on from the double Gaussian fits to [OIII] we then attempted to fit all the emission lines 
in the nuclear
regions with a double Gaussian model comprising two components:
\begin{enumerate}
\item [-] A narrow component (FWHM $=$ 383$\pm$15~km s$^{-1}$ ) with redshift $z =
  0.15225$, measured from double Gaussian fits to the bright low
  ionisation lines.
\item [-] An intermediate component (FWHM $=$ 1282$\pm$25~km s$^{-1}$ ) blue shifted by
    679$\pm$20~km s$^{-1}$ with respect to the narrow component, measured from double Gaussian fits to the 
    [OIII] lines.
\end{enumerate}
This model provides an adequate fit to all the lines in the nuclear emission spectrum, with 
the following notable exceptions:
\begin{itemize}
\item [-] The H$\alpha+$[NII] blend requires an additional broad H$\alpha$ emission line component  to 
fit the wings of the blend. We find that we can obtain an adequate fit to the blend if we set the 
width and redshift of the broadest H$\alpha$ component to be the same as those deduced from
single Gaussian fits to the broad Pa$\alpha$ line detected by Bellamy et al. (2003: $z= 0.15238$, 
FWHM $=$ 1940~km s$^{-1}$). The flux measured
for the broadest H$\alpha$ component is $4.2\times10^{-15}$ erg cm$^{-2}$ s$^{-1}$. The detection of 
this
component provides further evidence that the reddened quasar nucleus is detected at the longer wavelength end of the optical window.
\item [-] The faint high ionization lines [NeV]$\lambda$3425 and [FeVII]$\lambda$6087 could be fitted 
adequately by the broad component alone, whereas the faint low ionization
[NI]$\lambda$5200 could be fitted by the narrow component alone. Presumably, in each of these cases 
the other component that would otherwise be detected in a double Gaussian fit is not detected because 
of the low S/N and uncertainties in the continuum subtraction.
\end{itemize}

The velocity shifts and widths of the various kinematic components detected in PKS1549-79 are compared 
in Figure 5 and Table 2. Not suprisingly, the intermediate and narrow components in the double Gaussian 
fits have similar redshifts and line widths to the high and low ionization lines
measured in single Gaussian fits to the lower resolution data of Tadhunter et al. (2001). 
It is  
notable that both the width and velocity shift of the nuclear narrow component are consistent with those 
deduced for the extended H$\alpha$ emission
on both sides of the nucleus; the redshift of this component is also close to that measured for both 
the core of the broad Pa$\alpha$ line and the narrow HI absorption (see below). Therefore we identify 
the nuclear narrow component with the rest frame of the host galaxy, and the intermediate component
detected most clearly in the [OIII] lines with a highly disturbed component that is 
blueshifted 
relative to this rest frame.

\begin{figure}
\epsfig{file=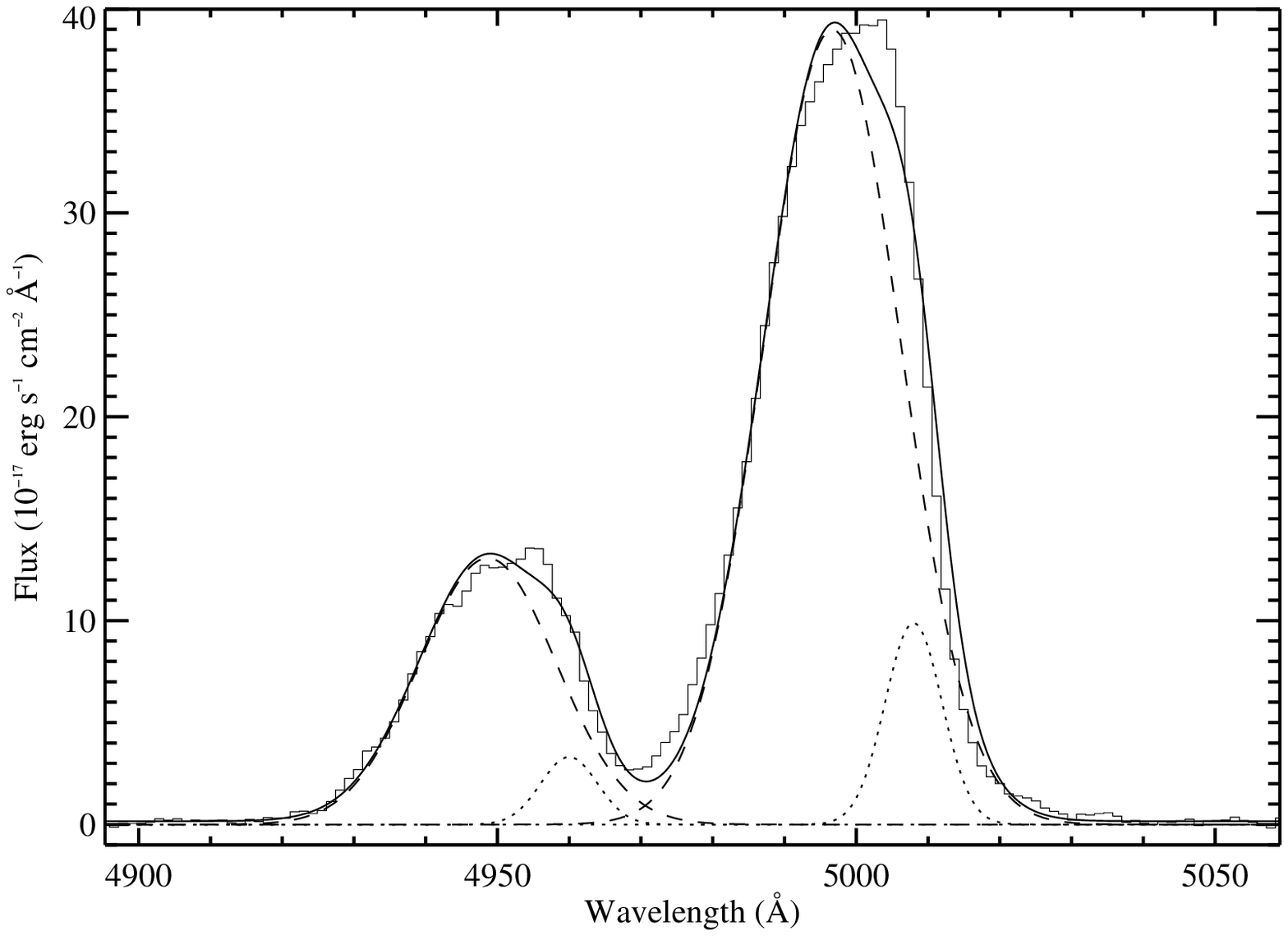,width=8.5cm,angle=0.}
\caption
{Double Gaussian model for the [O III]$\lambda\lambda$4959,5007 doublet in the nucleus,
forcing  the redshift of the narrow component to be consistent with
the redshift of the narrow component in the low ionisation lines. 
The composite model is overplotted (bold line) on the
extracted spectrum (faint line). All of the 4 components (2 for each line)
are also plotted: the dotted lines trace the
narrow components and the dashed lines trace the broad
components. }
\end{figure}

\begin{table}
\begin{tabular}{lll}
Component &Redshift &Line width(FWHM) \\
          &          &  (km s$^{-1}$)    \\
          \hline \\
Nuclear narrow &0.15225$\pm$0.00003 &383$\pm$15 \\
Nuclear intermediate  &0.14959$\pm$0.00003 &1282$\pm$25 \\
& & \\
Pa$\alpha$ core &0.15238$\pm$0.00006 &1250$\pm$70 \\
Pa$\alpha$ broadest &0.1529$\pm$0.0002 &3525$\pm$200 \\
& & \\
HI 21cm narrow  &0.15195$\pm$0.00006 &80$\pm$30\\
\end{tabular}
\caption{Comparison of the redshifts and line widths of the various kinematic 
components detected in the nuclear regions of PKS1549-79. Results for
nuclear narrow and intermediate components 
are based on double Gaussian fits to the optical emission lines detected  in the VLT
spectrum, while the Pa$\alpha$  results are based on double Gaussian fits to the Pa$\alpha$ 
feature detected in the new AAT K-band spectrum.} 
\end{table}

\subsubsection{Ionization and physical conditions}

Given the success of the two component emission line profile described in the last
section, it is possible to investigate the ionization, reddening and physical
conditions for the narrow and intermediate emission line components detected in the
nuclear aperture separately. The emission line ratios for both of the components
derived from the double Gaussian profile fits are shown in the second column of Table
3. In addition to the nuclear aperture, Table 3 also shows emission line ratios
determined for a $1.3\times1.3$ arcsecond aperture centred 1.6 arcseconds to the east of the
nucleus along PA75 (AP1: VLT spectrum).  In the case of this extended
aperture, which is affected by seeing disk
spillover of the nuclear emission line flux,  it proved necessary to use a double
Gaussian
fit to model the lines accurately -- one component representing
the nuclear
intermediate component and the other  the extended narrow component.
However, despite the contamination by the nuclear flux in this aperture, the narrow
component has a much better contrast relative to the intermediate component than in the
nuclear aperture; in Table 3 we only
list the narrow line ratios for this aperture.

It is immediately clear from Table 3 that the blueshifted, intermediate component 
detected in the nucleus is
characterised by a high ioniation state, with relatively strong [NeV], [NeIII],
HeII, [OIII] and [FeVII] lines, and relatively weak [OII], [OI] and [SII] lines. 

In contrast, the ionization state of the narrow component detected in a both the 
nuclear and extended apertures is
relatively low, with [OII], [OI], [NII] and [SII] all strong relative to the higher
ionization [NeV], [NeIII] and HeII and [OIII] lines.

On the basis of the H$\alpha$/H$\beta$ ratio alone there is no strong evidence for a high degree
of reddening affecting the narrow line region in this object, in contrast to the situation
in PKS1345+12 -- an object that also shows broad, blueshifted Balmer lines (Holt et al.
2003). However, it is
difficult to be entirely confident of the accuracy of the H$\alpha$ flux because of the
complexity of the 7 Gaussian model required to fit the H$\alpha$+[NII] blend. Moreover, the
higher order Balmer lines cannot be used to check the reddening estimated from 
H$\alpha$/H$\beta$ because of their weakness relative to the continuum and underlying
Balmer absorption line features. Given these uncertainties we have made no attempt to
correct the emission line ratios for reddening.

Similarly, it has not been possible to estimate accurate electron temperatures for the narrow
line gas using the [OIII](5007+4959)/4363 line ratio, because the faint [OIII]$\lambda$4363
line is highly sensitive to the accuracy of the continuum subtraction, particularly for the
broader, blueshifted component that is also blended with H$\gamma$. Nonetheless it has
been possible to estimate the electron density from the [SII](6731/6717) ratio
for the narrow emission line component detected in the nuclear aperture and the 
extended aperture along PA75. Using a model for the [SII]$\lambda\lambda$6717,6731 blend that
includes both an intermediate blueshifted component and a narrow component for each 
line in the doublet, and assuming an electron temperature of 10,000K for the region
emitting the [SII] lines, we estimate an electron density $430\pm50$~cm$^{-3}$ for the narrow 
component in the nuclear aperture. Unfortunately, 
because
of the relative weakness of the intermediate blueshifted component to the lines, the
electron density in the gas emitting the blueshifted component is not well-constrained by these
observations. We find that, although 4 Gaussian models with the doublet ratios left
unconstrained tend to the high density limit for the intermediate component, models with
the ratio of the intermediate component fixed at the low density limit also provide
adequate fits to the blend. On  the other hand, the ratios determined for the narrow components are not
sensitive to the details of the model used for the intermediate components.

\begin{figure*}
\begin{tabular}{cc}
\epsfig{file=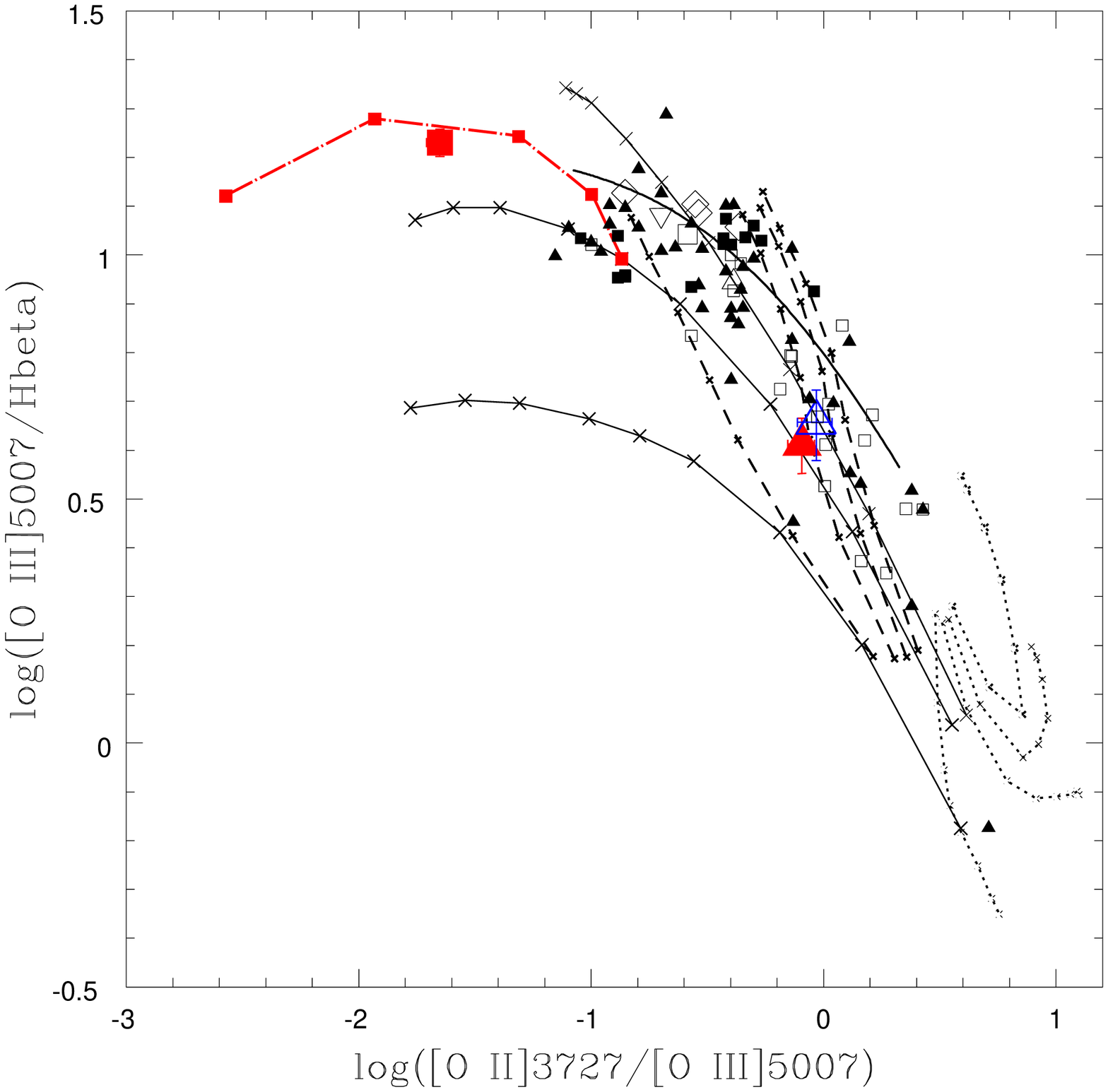,width=8.0cm,angle=0.} &
\epsfig{file=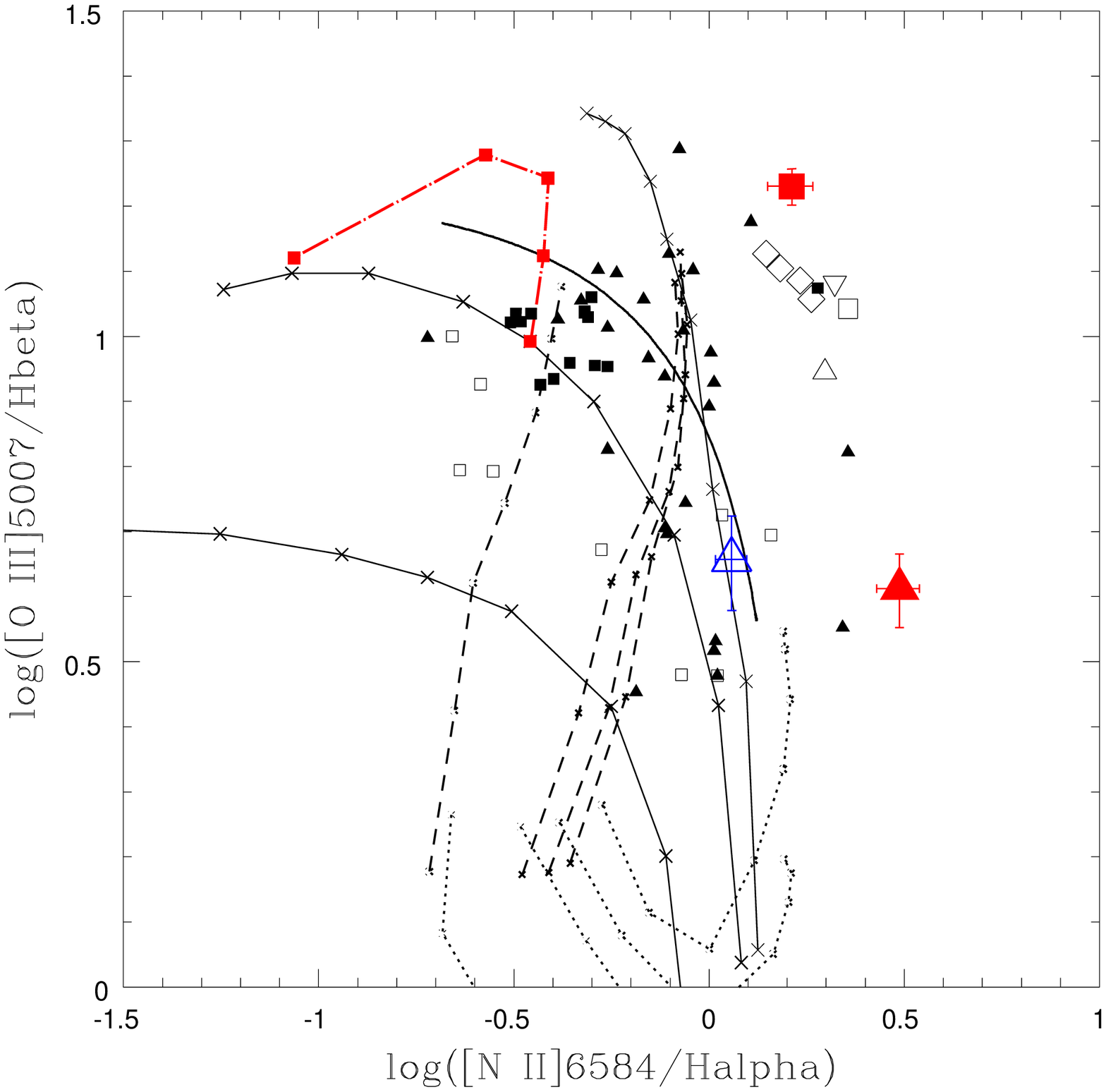,width=8.0cm,angle=0.} \\
\epsfig{file=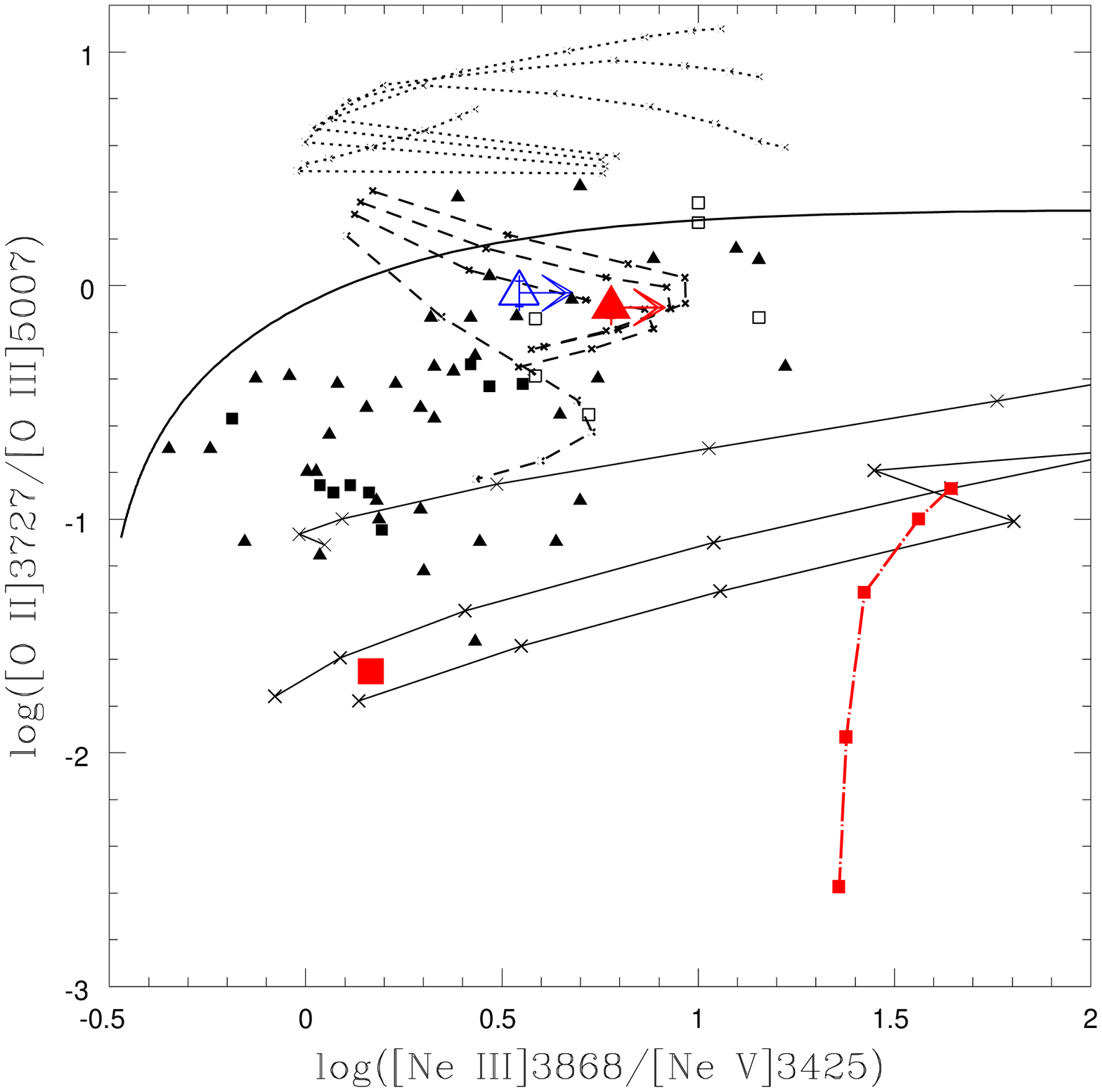,width=8.0cm,angle=0.} &
\epsfig{file=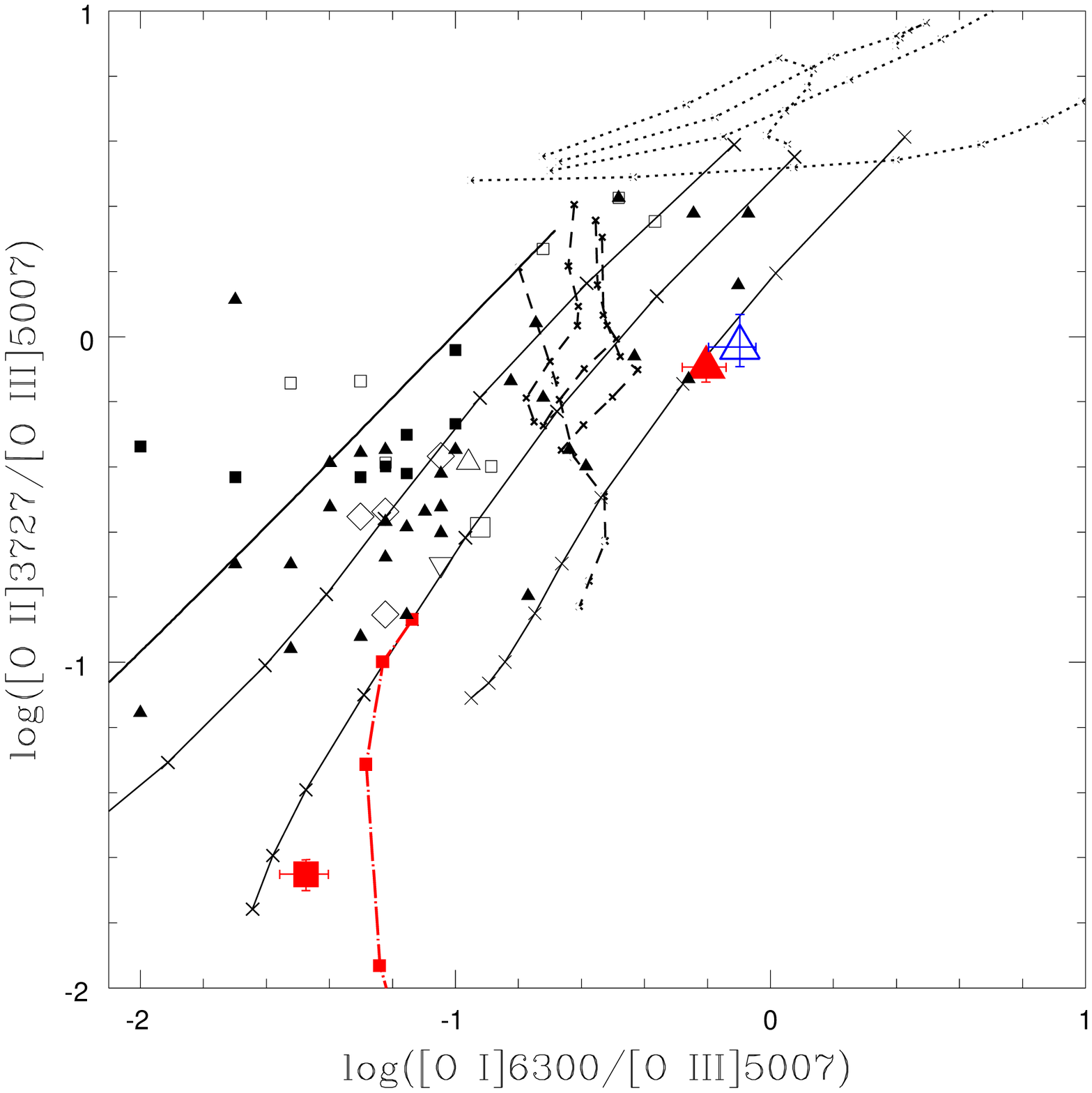,width=8.0cm,angle=0.} \\
\end{tabular}
\caption[Diagnostic diagrams for PKS 1549-79]{Emission line diagnostic diagrams for
  PKS 1549-79. {\it Upper left:} [OIII]/H$\beta$ vs. [OI]/[OIII]. {\it Upper right:}
  [OIII]/H$\beta$ vs.
  [NII]/H$\alpha$. {\it Bottom left:} [OII]/[OIII] vs. [NeV]/[NeIII]. {\it Bottom right:}
  [OIII]/[OII] vs. [OI]/[OIII]. In these diagrams, the pure shock models 
  of Dopita \& Sutherland (1996) for a range of magnetic parameter ($B/\sqrt{n} =$ 0, 1, 2
  $\mu$~G~cm$^{3/2}$) with shock velocities $150 \leq v_{shock} \leq 500$ km s$^{-1}$
  are represented by dotted lines, while the shock plus precursor models are represented
  by dashed lines. Solid lines represent optically thick power law ($F_{\nu} \propto
  \nu^{\alpha}$) photoionization models calculated using the MAPPINGS code for
  three power-law spectral indices ($\alpha =$ -1.0, -1.5, -2.0) for a sequence in the
  ionization parameter ($2.5\times10^{-4} \leq U \leq 10^{-1}$) increasing from
  right to left. Most of the 
  photoionization models plotted in these figures were calculated for low electron densities ($N = 10^2$
  cm$^{-3}$), however the dot dashed lines shows a sequence of models calculated at a single
  ionization parameter ($U = 5\times10^{-3}$) for a range of densitities ($10^2 < N < 10^6$ cm$^{-3}$).
 The bold line corresponds to mixed-medium models from Binette et al. (1996) 
  that include both matter-bounded and ionization-bounded clouds for a range of ratios of
  the solid angles covered by the matter bounded and ionization bounded clouds
  ($10^{-2} \leq A_{M/I} \leq 10$) increasing from right to left. The nuclear intermediate
  component is represented by the large solid square, while the narrow components for
 the nuclear and extended apertures are represented by large solid and open triangles
 respectively. For
  comparison, the small filled triangles represent data for the nuclear narrow line
  regions of radio galaxies (data from: Wills et al. 2002, Storchi-Bergmann et al. 1996, Tadhunter 1987,  Grandi \& Osterbrock 1978, Costero \& Osterbroack 1977,
  Cohen \& Osterbrock 1977, Cohen \& Osterbrock 1981, Dickson, 1997, Solorzano-Inarrea
  2001, Villar-Martin et al. 1998, Clark 1996, Robinson et al. 2000), 
  small filled squares represent extended emission line regions (data from: 
  Tadhunter et al. 1994, Storchi-Bergmann et al. 1996, Morganti et al. 1991, 
  Robinson et al. 2000), and open squares represent EELRs with strong jet-cloud interactions
  (data from: Villar-Martin et al. 1998, 1999, Clark 1996, Clarke et al. 1998, 
  Solarzano-Innarea 2001). On addition, we present the line ratios for various 
  kinematic components detected in the spatiallty resolved NLR of Cygnus A
  by Taylor, Tadhunter \& Robinson (2003), 
  with the nuclear broad and narrow components plotted as 
  small open triangles, and the extended narrow components plotted as open diamonds.
    }
\label{fig:diag1549}
\end{figure*}

In order to investigate the ionization mechanism(s) for the various
kinematic components, the emission line ratios are compared with various ionization models and the results for other
radio galaxies in the four diagnostic diagrams presented in Figure 7. Four sets of models 
are plotted: single slab optically thick power law photoionization models calculated
using the MAPPINGS code; mixed medium photoionization models taken from
Binette et al. (1996); pure shock cooling zone models 
and shock plus precursor models from Dopita \& Sutherland (1995, 1996).

These diagnostic diagrams further emphasise the high ionization character of the gas
emitting the intermediate, blueshifted component. In all diagnostic diagrams except
that involving [NII], the line ratios
for this component can only be fitted with power-law AGN photoionization models that have a
high ionization parameter ($0.05 < U < 0.1$), and the line ratios are most
consistent with models that have a power-law spectral index of -1.5. Although
the particular mixed medium photoionization models plotted in the diagrams do not provide
a good fit to the line ratios of the intermediate component, it is possible that, by changing the parameters of such models (e.g. ionization parameter, column depth of matter bounded
component), an acceptable fit could be obtained. It is also notable that none of the shock or shock+precursor can provide an adequate fit to the line ratios. 

The intermediate
component has line ratios that place it beyond the high ionization end of  the locus
of narrow line radio galaxies, overlapping with the range of ionization parameter
measured for the narrow, blueshifted absorption line systems detected against the UV and X-ray
continua of Seyfert 1 galaxies and quasars  (Crenshaw et al. 2003). Indeed we
would expect such absorption line systems to be detected in PKS1549-79 for
lines of sight to the AGN that intercept the high ionization gas. The only line ratio that does not fit in
with this general trend is [NII]/H$\alpha$, which is unusually large considering the
high ionization state revealed by the other lines. This may
reflect the fact that the systematic uncertainties for this ratio are larger than for
the others because of the complexity of the H$\alpha$+[NII] blend. Moreover, the models
plotted in Figure 7 assume Solar elemental abundance ratios, whereas the abundance 
of nitrogen relative to other elements such as oxygen may be
significantly larger than the Solar value -- the favoured explanation for the unusually
strong [NII] in, for example, Cygnus A (Tadhunter, Metz \& Robinson 1994).

The diagnostic diagrams also show that the line ratios of the narrow emission line
components (all apertures) fall at the lower ionization end of the ratios measured for the narrow line
regions of radio galaxies in general, consistent with photoionization at moderate ionization
parameter, or shock+precursor ionization models. Despite the evidence for young
stellar populations in this source, and the possibility of on-going star formation, the
strengths of the [OI] and [NII] emission lines relative to H$\alpha$ preclude
HII region-like photoionization by normal OB stars.

\begin{table}
\begin{center}
{\footnotesize
\begin{tabular}{lcrr}\\  \hline\hline
\multicolumn{1}{c}{Emission line} & \multicolumn{1}{c}{Comp.} 
 & 
\multicolumn{1}{c}{Nuclear}&\multicolumn{1}{c}{Extended} \\
\multicolumn{1}{c}{} & \multicolumn{1}{c}{}  
 & 
\multicolumn{1}{c}{flux} & \multicolumn{1}{c}{flux} \\
\multicolumn{1}{c}{$(a)$} & \multicolumn{1}{c}{$(b)$}  & \multicolumn{1}{c}{$(c)$} 
& \multicolumn{1}{c}{$(d)$} \\\hline
\\
{[Ne V]}  3425.9  & n                  & -      &$<20$            \\
            & i & 75 $\pm$ 8                    & -               \\
{[O II]}  3727.64$\dagger$ & n & 330 $\pm$ 30      & 420 $\pm$ 50  \\  
            & i          & 38 $\pm$ 8          & -               \\
{[Ne III]} 3868.8 & n & 50 $\pm$ 8              & 80 $\pm$ 10     \\
            & i & 110 $\pm$ 8                   & -               \\
{[Ne III]} 3968.4 & n  & 18 $\pm$ 3             & 26 $\pm$ 4      \\
            & i         & 38 $\pm$ 4            & -               \\
H$\gamma$  4339.95 & n  & 33 $\pm$ 4            & -               \\
            & i        & 27 $\pm$ 3             & -               \\
{[O III]}  4363.2 & n  & 8 $\pm$ 3              & $<$ 22          \\
            & i         & $<$ 8                 & -               \\
He II     4685.75  & n  & 11 $\pm$ 4            & 20 $\pm$ 10     \\
            & i         & 52 $\pm$ 7            & -               \\
H$\beta$   4860.75 & n  & 100                   & 100             \\
            & i         & 100                   & -               \\
{[O III]} 4958.9  & n   & 140 $\pm$ 10          & 150 $\pm$ 30    \\
            & i         & 580 $\pm$ 45          & -             \\
{[O III]} 5006.9  & n   & 410 $\pm$ 50          & 450 $\pm$ 80     \\
            & i            & 1700 $\pm$ 150     & -                \\
N I       5199.1  & n  & 51 $\pm$ 6             & -               \\
            & i              & $<$ 10           & -               \\
He I       5875.2 & n & 64 $\pm$ 8              & -               \\
            & i           & $<$ 10              & -               \\
{[Fe VII]} 6086.0 & n & $<$ 10                  & -                \\
            & i        & 42 $\pm$ 5             & -               \\
{[O I]}    6300.3 & n  & 160 $\pm$ 20           & 80 $\pm$ 10      \\
            & i        & 57 $\pm$ 10            & -                \\
{[O I]}    6363.8 & n & 53 $\pm$ 6              & 28 $\pm$ 4       \\
            & i          & 15 $\pm$ 4           & -               \\
{[N II]}   6548.09 & n & 420 $\pm$ 30           & 180 $\pm$ 20     \\
            & i          & 190 $\pm$ 20        & -                \\
H$\alpha$  6562.9 & n   & 400 $\pm$ 50          & 500 $\pm$ 70      \\
            & i           & 350 $\pm$ 40        & -              \\
{[N II]}  6583.36  & n & 1230 $\pm$ 100         & 560 $\pm$ 70     \\
            & i            & 570 $\pm$ 100      & -               \\
{[S II]}  6716.4  & n     & 290 $\pm$ 15        & 120 $\pm$ 20     \\
            & i            & 20 $\pm$ 4         & -                \\
{[S II]}  6730.8  & n     & 270 $\pm$ 15        & 140 $\pm$ 20     \\ 
            & i           & 47 $\pm$ 10         & -               \\
\\ \hline \\
F(H$\beta$) &n &$2.4\times10^{-16}$ &$0.9\times10^{-16}$ \\
(erg cm$^{-2}$ s$^{-1}$)  &i &5$.4\times10^{-16}$ & -  \\ 	    
\\\hline\hline
\end{tabular}
}
\caption
{Data for all optical emission lines detected in the nuclear and extended apertures of PKS
  1549-79.
Columns are: (a) emission line; (b)
emission line component where n and i are the narrow and intermediate
components;
 (c) emission line fluxes measured for a 1.3$\times$1.3 arcseond
 aperture centred on the nucleus, extracted from the PA75 VLT data (fluxes expressed
 relative to H$\beta$);
 (d) emission line fluxes measured for a 1.3$\times$1.3 arcseond
 extended aperture centred 1.6 arcseconds east of the
 nucleus, extracted from the PA75 VLT data.
\newline
$\dagger$ weighted mean of [O II] doublet at the low density
limit.}
\end{center}
\end{table}

\subsection{Near-IR spectroscopy}

Our new deep K-band spectrum of the nuclear regions
taken with the ATT/IRIS2 is shown in Figure 8. This represents a substantial improvement
over the previous K-band spectrum published in Bellamy et al. (2003), which was affected by a faulty
grism that resulted in a splitting of all spectral features. As well as the strong Pa$\alpha$
emission line, the spectrum shows a feature near 2.26$\mu$m that is likely to represent a
blend of H$_2\nu$=1-0S(3) and [SiVI]$\lambda$1.963$\mu$m, an  H$_2\nu$=1-0S(2)
line and (less certainly) a broad feature close to the expected wavelength of redshifted 
Br$\delta$.

\begin{figure}
\epsfig{file=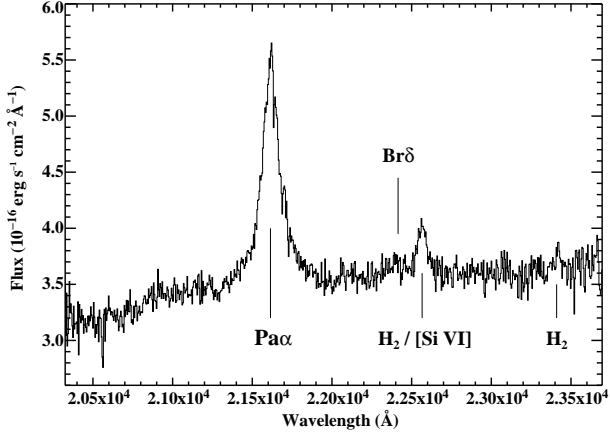,width=8cm, angle=0}
\caption{Near-IR (K-band) spectrum of PKS1549-79 taken using the IRIS2 spectrograph on the 
AAT. The main spectral features are marked. At near-IR wavelengths the spectrum is dominated by
a quasar component.}
\end{figure}

The Pa$\alpha$ feature can be modelled as a combination of two Gaussians of width 1220$\pm$90 
km s$^{-1}$ and 3540$\pm$320 km s$^{-1}$ (FWHM) with fluxes of $(1.10\pm0.16)\times10^{-14}$
erg cm$^{-2}$ s$^{-1}$ and $(1.92\pm0.15)\times10^{-14}$
erg cm$^{-2}$ s$^{-1}$ respectively; redshifts for these components
are given in Table 2. Similar results are obtained
from fitting our NTT/SOFI spectrum. 
Although the narrower of these two broad components
has a width similar to that of the intermediate component detected in the [OIII] lines, its
redshift is close to that of the narrow emission line components (see Table 2), therefore 
it cannot be emitted by
the same gaseous system that emits the intermediate [OIII] line component. Given this, 
and the fact
that the broader Pa$\alpha$ component has a width that falls outside the normal range for the
narrow line gas associated with AGN, it is likely that the Pa$\alpha$ line is dominated by
genuine broad line region (BLR) emission. For comparison, a single Gaussian fit to the
Pa$\alpha$ gives a linewidth: FWHM $=$ 1940$\pm$50~km s$^{-1}$. 

\begin{figure*}
\epsfig{file=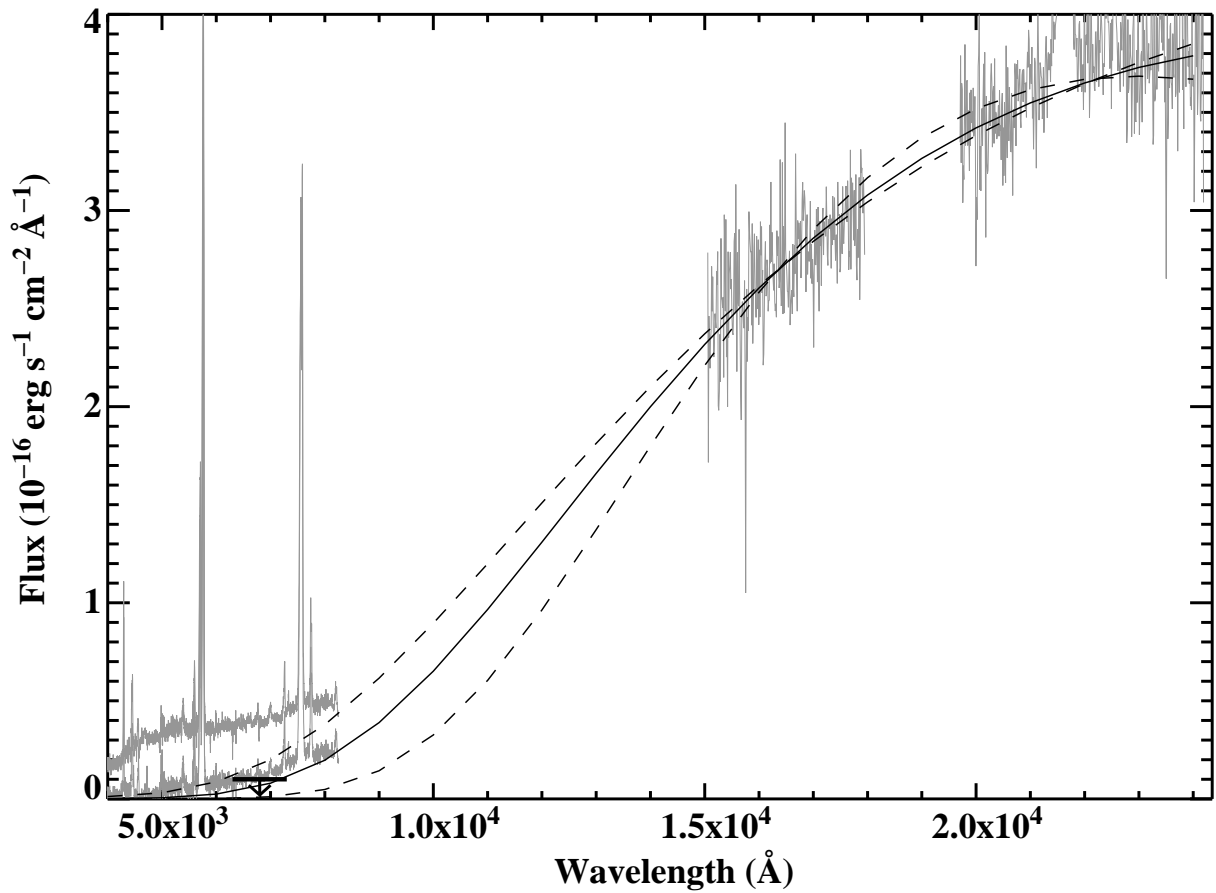,width=16cm, angle=0}
\caption{Fit to the shape of the near-IR H$+$K spectrum of PKS1549-79 taken on the NTT, showing
the extrapolation to optical wavelengths for various assumptions about the power-law shape of the optical-IR continuum.
Also shown are the VLT optical spectrum (upper curve at short wavlengths), and the VLT spectrum following subtraction of the best fitting
stellar continuum model (lower curve at short wavelengths). The upper and lower dashed lines represent
models with the minimum and maximum reddening respectively, while the solid line represents the
average reddening case (see Table 4). The arrow in the lower left corner of the plot 
represents the upper limit on the AGN continuum flux at 5900\AA\, derived from HST
imaging observations.}
\end{figure*}

The new spectra also allow us to resolve the uncertainty in the identification of the feature
at $\sim$2.26$\mu$m in the observed spectrum (see Figure 8). In terms of its redshift
this feature could either be identified with blueshifted [SiVI]$\lambda$1.963$\mu$m or with
H$_2\nu$=1-0S(3)
at the redshift of the nuclear narrow component. However, a single Gaussian fit to this feature
gives a width of 970$\pm$80~km s$^{-1}$, which is significantly broader than the nuclear narrow component, but
significantly narrower than the intermediate component, suggesting that the feature is
composite. Therefore we have attempted a double Gaussian fit to the feature, with one Gaussian
component representing H$_2\nu$=1-0S(3) with the same width and redshift as the nuclear 
narrow component, and the other representing [SiVI]$\lambda$1.963$\mu$m with the same width and
redshift as the nuclear intermediate component. This double Gaussian model provides a
good fit to the data with fluxes of  $(4.5\pm1.5)\times10^{-16}$~erg cm$^{-2}$ s$^{-1}$
and  $(3.1\pm0.3)\times10^{-15}$~erg cm$^{-2}$ for the narrow H$_2\nu$=1-0S(3) and
intermediate [SiVI]$\lambda$1.963$\mu$m components respectively. The detection of  H$_2\nu=1-0$~S(3)
is consistent with the presence of H$_2\nu$=1-0S(2) in the spectrum at longer wavelengths;
the estimated luminosity for the H$_2\nu$=1-0S(3) line ($2.3\times10^{40}$~erg s$^{-1}$) 
is less than estimated by Bellamy et al.
(2003) who assumed that all the $\sim$2.26$\mu$m feature is H$_2\nu=1-0$~S(3).
It is notable that the detection of the blueshifted [SiVI]$\lambda$1.963$\mu$m line is consistent 
with the high ionization character of the intermediate emission line component deduced from
the optical observations (see section 3.2).

Along with information about the emission line systems, the new spectra provide key
information about the quasar continuum emission. In particular, the new NTT spectrum has a
broader spectral coverage than the AAT/IRIS spectra and allows more stringent limits to be
placed on the reddening to the AGN. To estimate the AGN reddening we follow the
same technique as Bellamy et al. (2003): fitting the near-IR continuum spectrum assuming
a range of intrinsic power-law optical-infrared spectral shapes for quasars taken from 
Simpson \& Rawlings (2000).  The results are presented in Table 4 and Figure 9.

On the basis of this modelling alone it is clear that the V-band absorption and absolute
magnitude of the AGN must fall in the range $4.9 < A_v(intrinsic) < 13.2$
and $-27.56 < M_v < -22.64$ respectively. However, the recent HST/ACS imaging of PKS1549-79
(Batcheldor et al. 2006 in preparation)
allows us to place a stringent upper limit on the flux of the quasar core at 5900\AA (rest)
of $F_{\lambda} < 1.0\times10^{-17}$ erg cm$^{-2}$ \AA$^{-1}$. Using this limit it is clear 
from Figure 9
that the intrinsic absorption 
and luminosity of the quasar must lie at the upper end of the range predicted
on the basis of of the near-IR continuum modelling; we find that the
optical flux limit implies $A_v > 6.4$, $M_V < -23.5$ and $M_K <-27.3$. 
This places the AGN in PKS1549-79 firmly in the luminosity
range of quasars ($M_v < -23.0$). Note also that such high values of nuclear extinction and
luminosity are 
consistent with the residual spectrum obtained by subtracting the best fitting stellar continuum 
model from our VLT spectrum (see Figure 4).

\begin{table}
\begin{tabular}{lrrr}
\hline \\
 	       &Minimum	&Average &Maximum \\ \hline \\
$\alpha$(intrinsic) &1.62	&0.90	&-0.67 \\
E(B-V)	           &1.52	&2.31	&4.04 \\
A$_v$   &4.94	&7.52	&13.19 \\ \hline \\
M$_v$	            &-22.64    &-24.18	&-27.56 \\
M$_k$	           &-26.97    &-27.43	&-28.44 \\ \hline \\
\end{tabular}
\caption[]{Reddening and absolute magnitude estimates for the quasar nucleus in
PKS1549-79. The second, third and fourth columns give the minimum, average and maximum
reddening corresponding to the range of optical/UV power-law slopes deduced by Simpson \&
Rawlings (2000). The second row gives the values of the assumed spectral index for the
intrinsic optical-IR spectral shape (convention: $F_\nu \propto \nu^{-\alpha}$).
The final four rows give estimates for the E(B-V) reddening, A$_v$ extinction, and the estimated absolute magnitude 
of the quasar nucleus
in both the V and the K bands. In modelling the continuum shape we have assumed
the extinction law of Rieke \& Lebofsky (1985).}
\end{table}

\subsection{Radio imaging and spectroscopy}
\subsubsection{VLBI continuum imaging}

Figure 10 shows the VLBI images made with SHEVE at 2.3 and 8.4GHz. The radio source is
characterised by a compact core to the western extreme of the source, which shows a jet-like
extension in PA64, and a highly collimated jet structure to the east aligned along
PA120. The overall extent
of the source in the 2.3GHz image is 180mas (430~pc), and there is no strong evidence for
emission from more extended structures; $>$95\% of the total (single dish) flux is contained
within the structures visible in the VLBI images.
Relative to the western
core the jet appears significantly fainter in the 8.4GHz image (core/jet ratio $\sim$10:1) than it
does in the 2.3GHz image (core/jet ratio $\sim$2:1), but the western core has a similar
brightness in both images. The non-simultaneity of the 2.3GHz and 8.4GHz VLBI data, and the
resolution differences between the two frequencies, preclude accurate determination of the
power-law spectral indices of the jet and core components. However, the data are consistent
with a steep spectrum for the jet ($\alpha \sim -1.2$; $F_{\nu} \propto \nu^{+\alpha}$) and 
a much flatter spectrum ($\alpha \sim 0$) for the core. Note that the total flux recovered in
our VLBI observations at 8.4~GHz (3.6~Jy --- data taken in 1991) is significantly larger
than the 8.4~GHz flux measured more recently
by Drake et al. (2004b: 2.4~Jy --- data taken between 1996 and 1998). 
This suggests that the core is variable at high frequencies.

Overall, the radio structure is similar to the core-jet structures detected in some
other radio-loud quasars such as 3C273. Such structures are generally interpreted in 
terms radio emission from a two sided relativistic jet structure, with the axis of 
jets pointing close to our line of sight. In this case the radio emission from the jet on the
near side
is enhanced because the radio emission is beamed towards the observer, whereas the emission
from the jet on the far side is dimmed because the radiation is beamed away from the
observer. The significant curvature in the jet is also consistent with the beaming hypothesis, since
a small degree of intrinsic jet curvature will be exaggerated if the jet is pointing
close to the line of sight. If this
beaming/orientation explanation for the one-sided jet is correct, then the total radio
emission from the source will be boosted relative to the situation in which the jet
is close to the plane of the sky. Therefore, although the total radio power of PKS1549-79
uncorrected for beaming ($P_{2.3GHz} = 5.4\times10^{26}$~W Hz$^{-1}$) would lead to its classification as a powerful radio galaxy that
would normally appear as an FRII radio source if orientated close to the plane of the sky,
the true (unbeamed) radio power of the
source could be much lower. If we assume that the radio flux not recovered in the VLBI maps 
(i.e. 5\% of the total) represents an unbeamed radio component, this leads to a lower
limit on the total unbeamed radio power of $P_{2.3GHz} = 2.7\times10^{25}$~W Hz$^{-1}$ -- 
just above the break in the radio luminosity function (Auriemma et al. 1977) and in 
the transition region between FRI and FRII sources (Fanaroff \& Riley 1974). 

Assuming that the one sided nature of the jet is a consequence of beaming effects, we can use the
lower limit on the jet to counter-jet ratio ($R_{jc}$) derived from the radio maps to estimate the 
inclination of the jet to
the line of sight. The 21cm LBA continuum map  (see Figure 11) is the most sensitive
for the detected of extended jet features close to the nucleus. The peak flux in the jet detected in this
map is 2.3 Jy beam$^{-1}$, whereas the 3$\sigma$ detection limit for the counter jet is 48 mJy beam$^{-1}$.
Therefore a lower limit on the jet to counter-jet ratio is $R_{jc} \geq 48$. The jet to counter-jet ratio is
related to the bulk Lorentz factor in the jet ($\gamma$) and the inclination of the jet to the line of
sight (i) by:
\begin{equation}
cos(i) = \left( \frac{\gamma^2}{\gamma^2 -1} \right)^{1/2} \left( 
\frac{R_{jc}^{1/(2-\alpha)} -1}{R_{jc}^{1/(2-\alpha)}+1} \right)
\end{equation}
where $\alpha$ is the spectral index of the jet continuum emission. For $\alpha = -1$  and $R_{jc} \geq 48$ we estimate an upper limit on the inclination of $i < 55$ degrees, assuming that the Lorentz factor for the
bulk motion of the jet falls within the typical range estimated for extragalactic radio jets ($5 < \gamma < 10$).

An alternative possibility is that the one-sided 
jet is a consequence of a stronger interaction between the radio jet and the interstellar
medium on the east side of the nucleus that leads to local re-acceleration of the electrons
and enhanced radio emission; a strong jet-cloud interaction would also explain
the curvature in the jet. However, this scenario would not account for the relative strength of the
flat spectrum core and the fact that the asymmetric character of jet extends into the near-jet close to the core, 
as is clear from the higher resolution 2.3GHz map (Figure 10 middle).

\begin{figure}
\epsfig{file=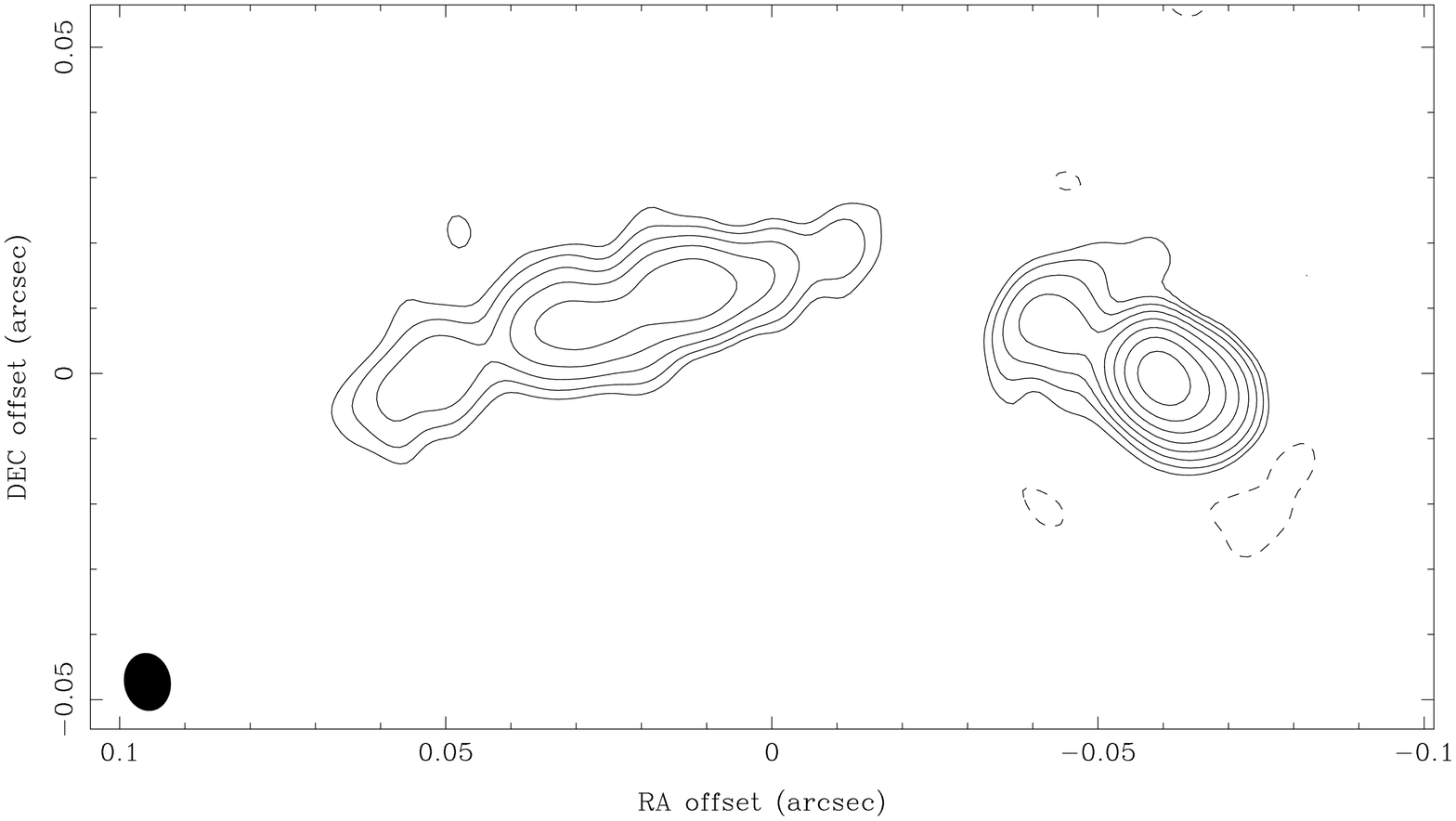,width=4.5cm,angle=-90}
\epsfig{file=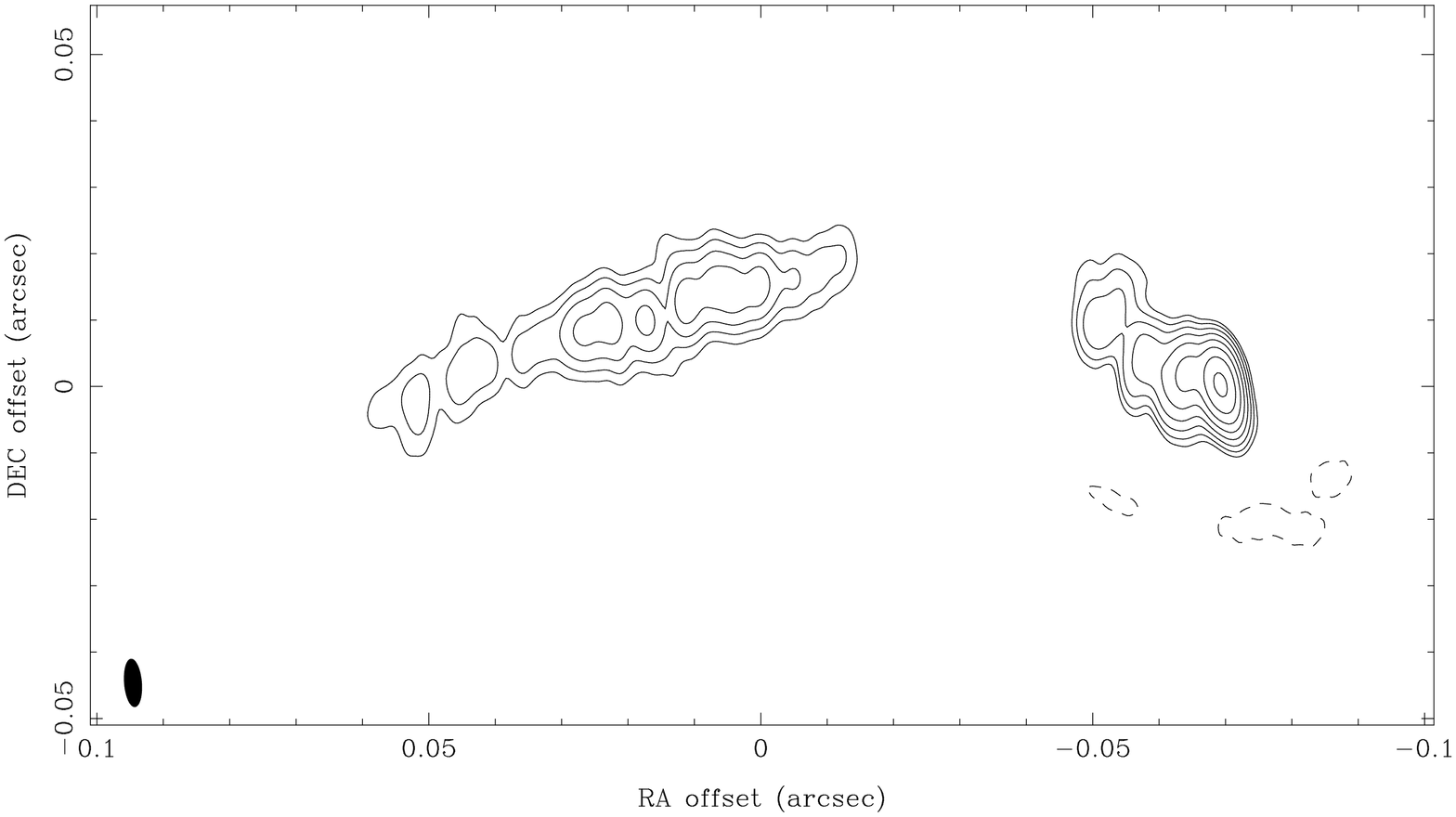,width=4.5cm,angle=-90}
\epsfig{file=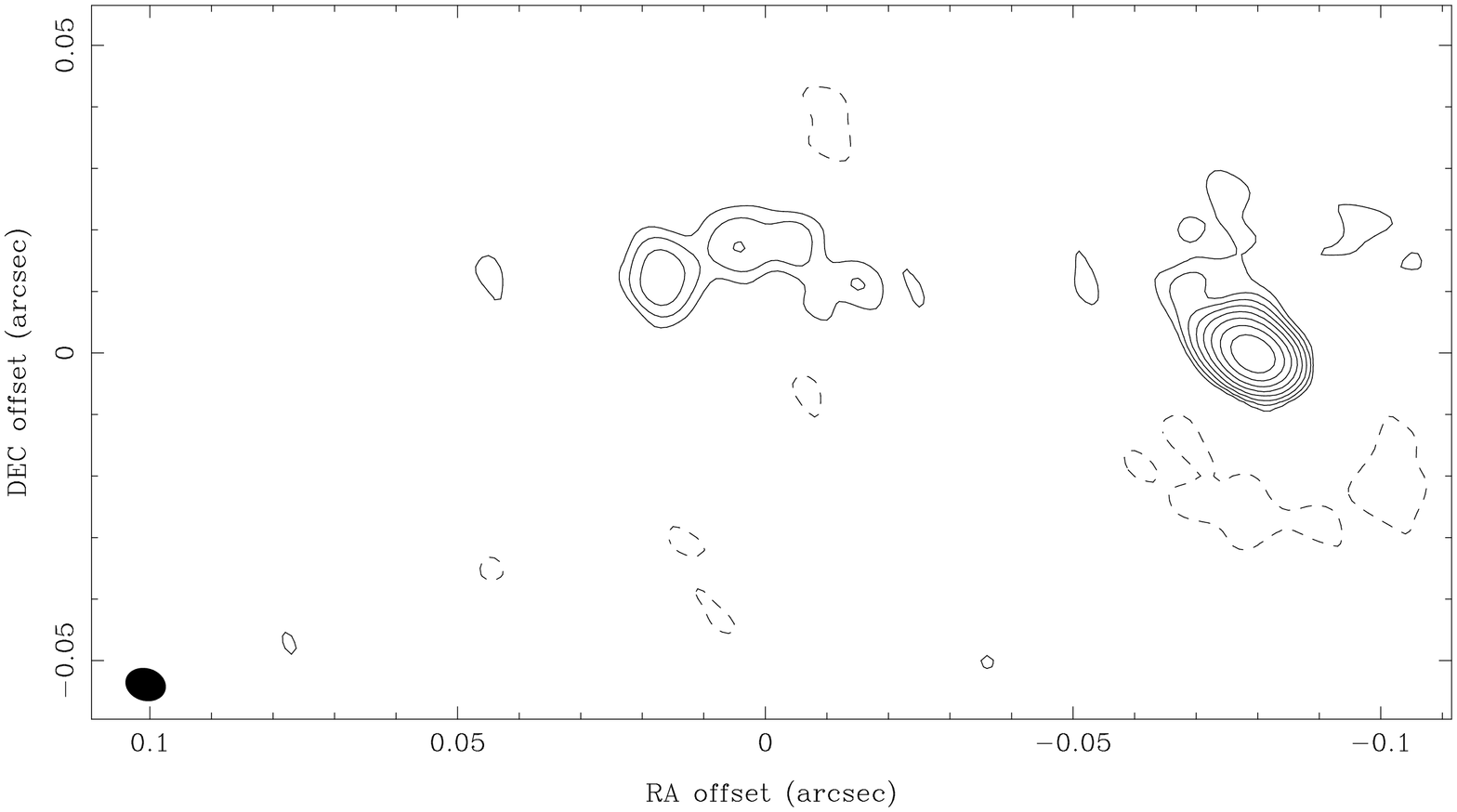,width=4.5cm,angle=-90}
\caption{High resolution VLBI radio maps of PKS1549-79. The upper map is for a frequency
of 2.3GHz  excluding the Hartebeesthoek antenna, the
middle map is also for a frequncy of 2.3~GHz  but
including the Hartebeesthoek antenna, 
while the lower map is for an observation
frequency of 8.4~GHz (contours: -10,10,20,40,80,160,320,640,1280 mJy/Beam).}
\end{figure}

\subsubsection{HI 21cm kinematics}

Given the core-jet character of the radio continuum structure, which suggested that
the jet is pointing close to our line of sight, it was a surprise to detect HI
absorption in our ATCA observations (Morganti et al. 2001). On the basis of
the simplest versions of the unified schemes one would normally expect a relatively unobscured view of the quasar nucleus from the direction of the radio jet (Barthel 1989). The absorption in the spatially integrated radio emission has a FWHM$\sim$80 km s$^{-1}$ and the peak optical depth is about 2\%. This leads to a column density of $(4.0\pm 0.5)\times10^{18}T_{spin}$ cm$^{-2}$, where $T_{spin}$ is the HI spin temperature. A possible interpretation described in Tadhunter et al. (2001) is that the HI absorption is associated with a dusty
cocoon around the nucleus that also emits the narrow emission line component, and is responsible for the obscuration of the quasar nucleus.

The new LBA data provide information about the spatial structure of the HI absorption. Figure 11 shows the profiles of the HI absorption measured at the locations
of both the core and the jet structures. Consistent with the idea of a foreground absorbing screen proposed by Morganti et al. (2001) and Tadhunter et al. (2001), the HI absorption was detected against both of the
major radio components with similar redshift, linewidth and optical depth. Therefore the diameter of the HI absorbing region
is $>120$ milliarcseconds ($> 286$~pc). Note that the narrow HI absorption is significantly blueshifted (by $\sim$75~km s$^{-1}$) relative to the narrow optical emission lines (see Figure 4
and Table 2), but this blueshift is an order of magnitude smaller than that measured in the [OIII] lines, and might be attributed to gravitational motions if the absorbing cloud has not yet settled into the 
circumnuclear disk.

In Figure 12 the HI absorption profile from the ATCA observations (from Morganti et al. 2001) is superimposed on the integrated absorption profile obtained from the new LBA observations. This shows that most of the HI absorption is recovered in the high spatial resolution LBA observations, and that the HI kinematics in the LBA and ATCA are consistent.

\begin{figure*}
  \centerline{\epsfig{figure=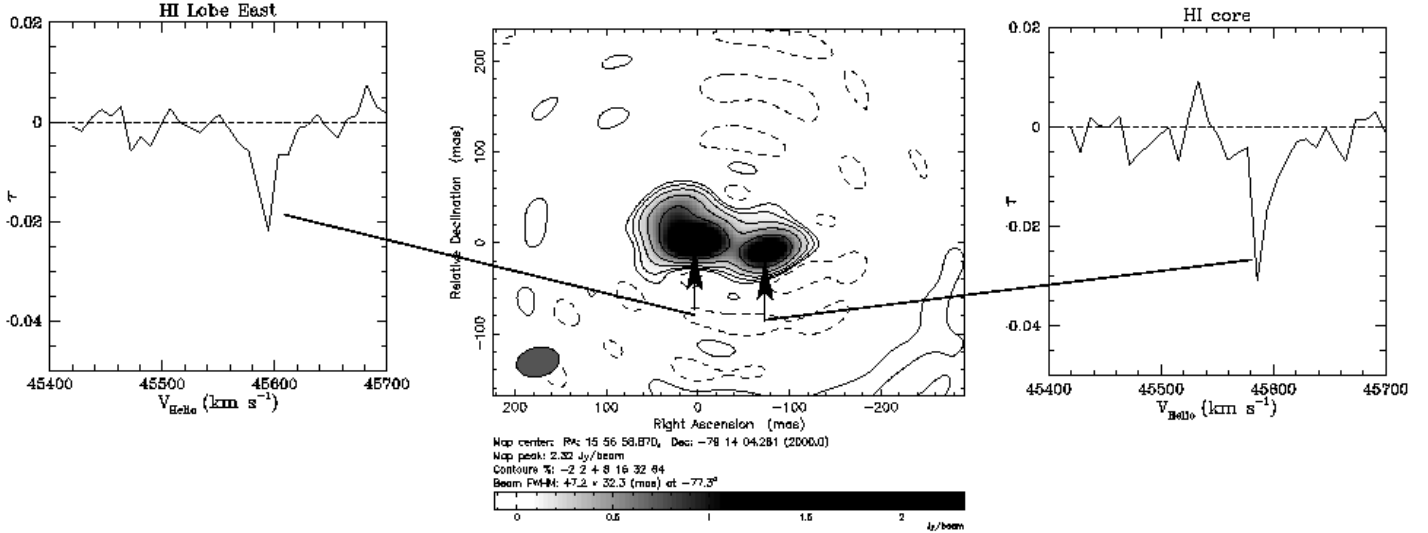,width=7cm,angle=270}} \caption{{\it
      Middle:} 1.64~GHz LBA continuum image of PKS1549-79 obtained from the
    line-free channels. {\it Right:} integrated profile of
    the HI absorption in the western core source.{\it Left:}: integrated
    profile of the HI absorption for the eastern jet source.   } 
\end{figure*}

\section{Discussion}

\subsection{Evolutionary status and the triggering of the activity}
The ready detection of multiple tidal tails at a relatively high surface brightness level is entirely consistent with the idea that the activity in PKS1549-79 has been triggered in a major merger between galaxies, at least one of which was gas rich and had a cold disk component prior to the merger. Indeed, the tail structure observed in PKS1549-79, with two near-parallel tails pointing to the south and a jet-like structure pointing to the north bears marked similarity with the structures produced in numerical simulations by Barnes \& Hernquist (1996) of the merger between two disk galaxies, observed from the orbital plane of the merging galaxies. In such a merger, it will take of order 500~Myr for the nuclei of the two galaxies to finally merge together. Therefore it is likely that that we are observing the system a considerable period after the start of the encounter.

The young stellar populations (YSP) provide useful complementary information about the merger. The presence of a reddened YSP in the nuclear regions of this object is consistent with the idea that the merger also triggered a starburst. Given that the YSP has an estimated age of 50 -- 250~Myr at the observation epoch, it is likely that the starburst was triggered a considerable period after the start of the merger sequence. This in turn suggests that one of the galaxies had a significant bulge that hindered the formation of a bar that would otherwise lead to the major merger-induced star formation occuring early in the merger (Mihos \& Hearnquist 1996). Nothwithstanding the uncertainties about the heating mechanism for the dust emitting the far-IR radiation, the substantial far-IR excess in this source suggests the presence of substantial ongoing star formation in the nuclear regions that is simultaneous with the currently observed quasar and jet activity.

It is also interesting to consider at what stage the radio source was triggered in the merger sequence. Tadhunter et al. (2001) have argued that, since the extreme forbidden emission line widths measured in PKS1549-79 are only detected in radio galaxies with young, compact radio sources, it is likely that we are observing the radio source at a relatively early stage after the triggering of the activity ($< 10^6$yr). If this is correct, it is plausible that there was a significant delay ($>$40 Myr) between the starburst associated with the YSP detected at optical wavelengths and the triggering of the radio jet activity. However, the evidence for such a time delay is weaker in PKS1549-79 than for some other radio sources in which the time lag between the the major episode of merger-induced star formation and the triggering of the currently observed radio source is estimated as 0.3 -- 2.0 Gyr
(Tadhunter et al. 2005, Emonts et al. 2005, Johnston et al. 2005). 

\begin{figure}
  \centerline{\epsfig{figure=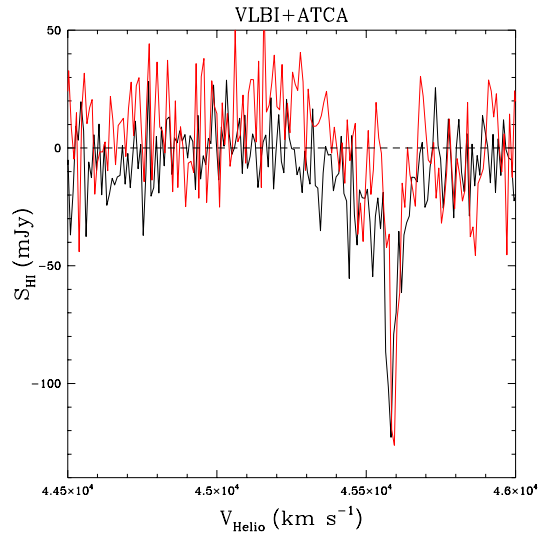,width=7cm,angle=0}} 
\caption{Profile of
  the HI absorption in PKS~1549--79. The solid line represents
  the ATCA data of Morganti et al. (2001), while the dotted lines represent the integrated HI profile 
  measured from the new LBA data. Most of the absorbed flux
    is recovered in the higher resolution LBA observations. }
\end{figure}

\subsection{The nature of the AGN}

Our new data provide key information about the nature of the AGN in PKS1549-79.
Most notably, the relatively narrow Pa$\alpha$ detected in the quasar component of PKS1549-79 (FWHM $=$ 1940$\pm$50~km s$^{-1}$ from a single Gaussian fit) is consistent with the classification of this source as a narrow line Seyfert 1 galaxy 
(NLS1: FWHM $<$ 2000~km s$^{-1}$: Osterbrock \& Pogge 1985, Osterbrock 1987) or more accurately, given its luminosity, a narrow line quasar. Interestingly, the extreme blueshift of the [OIII] line relative to the galaxy rest frame
in this object is also similar to those detected in the broad [OIII] emission line components of many NLS1 galaxies and similar objects 
(Veron-Cetty \& Veron 2001, Zamanov et al. 2002, Boroson 2005, Aoki, Kawaguchi\& Ohta 2005).

NLS1 are often interpreted as AGN that are accreting at a relatively high Eddington radio ($L_{bol}/L_{edd} \sim 1$), based on the small virial black hole mass estimates deduced from their relatively narrow permitted line widths
(Pounds, Done \& Osborne 1995, Mathur 2000, Mathur \& Grupe 2005). It has also been suggested that the NLS1 fall significantly below the correlations between black hole mass and host galaxy bulge properties (Grupe \& Mathur 2004). 
However, it is important to add the caveats that (a) the small permitted line widths of NLS1 may reflect orientation effects (e.g. BLR disk viewed face on) rather than low black hole masses; and 
(b) the bulge properties for NLS1 in the Grupe \& Mathur (2004) study were deduced from [OIII] linewidths, which are likely to be affected by non-gravitational (radial) motions suggested by the large [OIII] blueshifts detected in some sources. 

It is interesting to consider whether PKS1549-79 is also a high accretion rate object with a large Eddington ratio. Its black hole mass can be estimated using two independent methods. First, a virial mass estimate can be obtained using the Pa$\alpha$ line width and luminosity. Assuming the mimimum extinction for Pa$\alpha$ ($A_k = 0.74$ magnitudes), and a Pa$\alpha$/H$\beta$ ratio set at the case B recombination value (Pa$\alpha$/H$\beta = 0.332$, see Osterbrock 1989), we deduce an intrinsic quasar H$\beta$ luminosity of $L(H\beta) = 1.0\times10^{43}$~erg s$^{-1}$ for our adopted cosmology. Assuming that FWHM(H$\beta$) $=$ FWHM(Pa$\alpha$) $=$ 1940~km s$^{-1}$ we then use the relationship between FWHM(H$\beta$), L(H$\beta)$ and M$_{bh}$ published by Greene \& Ho (2005: their eqn 7) to estimate a virial black hole mass of $M_{bh} = 5.3\times10^7$~M$_{\odot}$. A second black hole mass estimate can be obtained from the correlation between R-band bulge luminosity ($M_R$) and black hole mass published by McClure and Dunlop (2002) along with the published R-band absolute magnitude for PKS1549-79 (Drake et al. 2004a: $M_R = -21.9$ for our cosmology)\footnote{Note that the R-band absolute magnitude derived by Drake et al. (2004a) is based
in the integrated light measured within the outermost ellipse fitted to their R-band image (corresponding
to a semi-major axis of 12.5 arcseonds or 30~kpc). Since this magnitude corresponds to the total light,
and no disk/bulge separation has been attempted, it is likely to {\it overestimate} the  
luminosity of the bulge.}. This leads to $M_{bh} = 2.6\times10^8$~M$_{\odot}$ --- almost an order
of magnitude larger than the virial estimate of black hole mass. However, the R-band luminosity estimate can be regarded as an upper limit because the YSP in the bulge and tail features of PKS1549-79 along with any AGN contamination are likely to artificially boost the R-band luminosity; the only way in which the R-band magnitude can substantially underestimate the black hole mass is if the host galaxy as a whole is subject to substantial intrinsic extinction. In contrast, the virial estimate is likely to represent a lower limit on the black hole mass because of the effects of orientation on the Pa$\alpha$ line width, especially given that the radio observations suggest that radio jet is pointing close to the line of sight, with the putative BLR disk observed close to face-on.  The true mass of the black hole in PKS1549-79 is therefore likely to fall between the virial and bulge luminosity estimates.

The other element required to determine the Eddington ratio is the bolometric luminosity ($L_{bol}$) of the quasar. $L_{bol}$ can be estimated using the quasar continuum modelling results along with information on generic quasar spectral energy distributions (SEDs) presented in Elvis et al. (1994).   
The range of optical SEDs that are consistent with our optical/IR spectroscopy and imaging have monchromatic V-band luminosities in the range $1.2\times10^{41} < L_V < 4.8\times10^{42}$ erg s$^{-1}$ \AA$^{-1}$ or $6.6\times10^{44} < \lambda L_{\lambda} < 2.6\times10^{46}$ erg s$^{-1}$ ($\lambda =5500$\AA) for the quasar in PKS1549-79. The V-band bolometric correction factor listed in Elvis et al. (1994) is: $L_{bol}/(\lambda L_V)=14.2$. Therefore, we estimate that the bolometric luminosity of PKS1549-79 falls in the range $9.0\times10^{46} < L_{bol} < 3.8\times10^{47}$ erg s$^{-1}$. Note that, although the radio jet in this source is pointing close to our line of sight, there is no evidence that the optical/IR continuum is substantially affected by contamination by a beamed synchrotron source: not only does the spectral energy distribution of PKS1549-79 show a steep decline between radio and sub-mm  wavelengths (compare Drake et al. 2004b with Beasley et al. 1997), but both the H$\beta$ equivalent width and the $L_{bol}/L_{ir}$ ratio fall well within the range measured for un-beamed quasars. 

These estimates of $L_{bol}$ lead to Eddington ratios in the range $0.3 < L_{bol}/L_{edd} < 11$ for our upper estimate of the black hole mass ($M_{bh} = 2.6\times10^8$~M$_{\odot}$) and in the range $1.3 < L_{bol}/L_{edd} < 57$ for our lower estimate of the black hole mass ($M_{bh} = 5.3\times10^7$~M$_{\odot}$). The measurement of such high Eddington ratios reinforces the link between PKS1549-79 and NLS1 galaxies.

Although we cannot rule out the idea that PKS1549-79 is a significant outlier relative to the statistical correlations we have used to derive both black hole mass and bolometric luminosity estimates, our results are at least consistent with accretion at a high Eddington ratio in this object. On this evidence,
accretion at high Eddington ratio does not prevent the formation of powerful, relativistic jets.
For comparison, the Eddington ratio measured for Cygnus A -- the only quasar with a direct, dynamical black hole mass estimate -- is ($1.6$ -- $6.0$)$\times10^{-2}$ (Tadhunter et al. 2003). Similarly, Dunlop et al. (2003) derive relatively small Eddington ratios (typically $L_{bol}/L_{edd} < 0.1$) based on R-band black hole mass estimates of several radio-loud and radio-quiet quasars. On the
other hand, Punsly \& Tingay (2005) have recently highlighted the high redshift quasar PKS0743-67 as 
another object that is accreting at high Eddington ratio but has powerful radio jets. 

\subsection{Characterising the warm outflow}

Recent theoretical studies of galaxy evolution have invoked quasar-induced outflows to explain both the correlations between the masses of black holes and the galaxy bulge properties (e.g. Silk \& Rees 1998, Fabian 1999, di Matteo et al. 2005), and the form of the galaxy luminosity function at the high luminosity
end (e.g. Benson et al. 2003). Such feedback models require that a relatively large fraction of the available accretion power of the quasars is thermally coupled to the circumnuclear ISM ($\ge$10\%: Fabian 1989, di Matteo et al. 2001).  

PKS1549-79 appears to represent a situation close to that modelled in some of the recent simulations (e.g. di Matteo et al. 2005), in which the black hole is growing rapidly via merger-induced accretion following a merger between gas rich galaxies. We have also found -- in the blueshifted [OIII] lines -- clear evidence for an outflow in the central regions of this source. Therefore it is important to investigate whether the properties of the warm gas outflow in PKS1549-79 are consistent with the assumptions of the quasar feedback models.

We start by determining the mass outflow rate in the intermediate (blue shifted) component. For a steady state outflow the mass outflow rate is given by:
\begin{equation}
\dot{M} = N \epsilon v_{out} A m_p
\end{equation}
where $N$ is the electron density, $\epsilon$ the volume filling factor, $v_{out}$ the outflow velocity, $A$ the area of the outflow and $m_p$ the mass of the proton. The filling factor is related to the H$\beta$ luminosity ($L(H\beta)$), density and emitting volume $V$ as follows:
\begin{equation}
\epsilon = \frac{L(H\beta)}{\alpha_{H\beta}^{eff} h \nu _{H\beta} N V}
\end{equation}
where $\alpha_{H\beta}^{eff}$ is the effective H$\beta$ recombination coefficient (see Osterbrock 1989), $\nu_{H\beta}$ the frequency of H$\beta$, and $h$ the Planck constant. It follows that the mass outflow rate can be written as:
\begin{equation}
\dot{M} = \frac{L(H\beta) m_p v_{out} A}{\alpha_{H\beta}^{eff} h \nu _{H\beta} N V}
\end{equation}
and for a spherical outflow this reduces to:
\begin{equation}
\dot{M} = \frac{3 L(H\beta) m_p v_{out} }{N \alpha_{H\beta}^{eff} h \nu _{H\beta} r}
\end{equation}
where r is the radius of the spherical volume. On the basis of our spectroscopy we estimate an H$\beta$ luminosity of $L(H\beta) = 2.8\times10^{40}$ erg s$^{-1}$ for the intermediate, blueshifted component and a rest frame outflow velocity of $v_{out} = 680$ km s$^{-1}$. If we assume case B recombination theory for an electron temperature $T = 10,000$~K and that the spherical outflow has the same radius as the radio source (180 milli-arcseconds or $r = 430$pc for our cosmology) we find that the mass outflow rate 
in units of solar masses per year is given by:
\begin{equation}
\dot{M} = 9.1 \left( \frac{10^2 cm^{-3}}{N} \right)  \; M_{\odot}\: yr^{-1}.
\end{equation}
The largest uncertainty in this calculation is the electron density of the outflowing gas ($N$). Because it has not been possible to measure the electron density for the outflowing gas directly in the case of PKS1549-79, we assume that it falls in the range measured for the NLR of other radio galaxies. The typical electron densities measured for kpc-scale ionized gas in extended radio sources such as Cygnus A (Taylor, Tadhunter \& Robinson 2003) and 3C321 (Robinson et al. 2000) are $N \sim 10^2$ cm$^{-3}$. However, the densities may be larger in the emission line regions associated with compact radio sources, and in the case of PKS1345+12 -- a similar object that has broad, blueshifted [OIII] lines -- a relatively high electron density ($N > 5\times10^3$~cm$^{-3}$) has been deduced for the blueshifted emission line components (Holt et al. 2003). Therefore it is reasonable to assume densities in the range $10^2 < N < 10^4$~cm$^{-3}$, leading to mass outflow rates $0.12 < \dot{M} < 12$~M$_{\odot}$ yr$^{-1}$. It is notable that this covers same range as the mass outflow rates deduced for the NLR of Seyfert galaxies (Veilleux, Cecil \& Bland-Hawthorn 2005) and the high ionization absorption line systems detected at UV and X-ray wavelengths in some AGN (Crenshaw et al. 2003), but is somewhat less than deduced for the neutral outflows detected in extreme starburst sources such as ULIRGs (Rupke et al. 2005a,b). For the same range of densities we estimate that the total mass in the warm gas outflow falls in the range $1.9\times10^4 < M_{total} < 1.9\times10^6$~M$_{\odot}$ and the filling factor falls in the range $2.3\times10^{-3} < \epsilon < 2.3\times10^{-7}$.

From the point of view of gauging the likely impact of the warm gas outflow on the circumnuclear gas in the bulge of the host galaxy of PKS1549-79 it is also important to estimate the kinetic power of the outflow, including both the radial and turbulent components in the gas motion. Assuming that the relatively large linewdith of the outflowing gas reflects a turbulent motion that is present at all locations in the outflow region, the kinetic power is:
\begin{equation}
\dot{E} = 6.34\times10^{35}\frac{\dot{M}}{2}(v_{out}^2 + (FWHM)^2/5.55) \;\; erg\: s^{-1}
\end{equation}
where $FWHM$ is the full width at half maximum of the blueshifted [OIII] line in km s$^{-1}$,
and $\dot{M}$ is the mass outflow rate expressed in solar masses per year. For $v_{out} = 680$ km s$^{-1}$, $FWHM = 1282$ km s$^{-1}$ and the mass outflow rates determined above, we find that the kinetic power for the outflowing gas falls in the range $2.2\times10^{40} < \dot{E} < 2.2\times10^{42}$~erg s$^{-1}$. Taking the Eddington luminosity derived from the larger black hole mass estimate (see section 4.2 above) as an estimate of the total power available from accretion, the kinetic power in the warm gas component represents only a small fraction ($7\times10^{-7} < \dot{E}/L_{edd} < 7\times10^{-5}$) of the available accretion power, in contrast to the quasar feedback models which require a much larger fraction of the accretion power of the black hole to be thermally coupled to the gas surrounding the black hole. It is also unlikely that the warm gas outflow by itself is capable of removing all the warm/cool gas from the central regions of the host galaxy. Assuming that PKS1549-79 has a total mass $M_{total} \sim 10^{11}$~M$_{\odot}$, and  a gas mass $M_{gas} \sim 10^{10}$~M$_{\odot}$ contained within a radius of 5~kpc -- conservative assumptions for ULIRGs which are likely to have a larger gas mass concentrated within a smaller radius -- the gravitational binding energy of the gas is $E_{bind} \approx G M_{gas} M_{total}/R_{gas} \approx 2\times10^{58}$~erg. In comparison,  the warm gas outflow will deposit only $\sim 10^{55}$ -- $10^{57}$~erg into the surrounding ISM, assuming that it can persist in its current form for the typical $10^7$~yr lifetime of an extragalactic radio source.

How do we explain the fact that the estimated kinematic power in the warm gas outflow in PKS1549-79 is at least three orders of magnitude less than might be expected on the basis of the quasar feedback models? Considering first the assumptions that have gone into the above estimates, the key parameters are the radius of the outflow, the electron density and the emission line luminosity. $\dot{E}$ would be 
underestimated if the radius and/or density of the outflow have been overestimated, or the H$\beta$ luminosity has been underestimated. However, it is unlikely that the radius of the outflow region has been overestimated, because recent HST imaging observations of PKS1549-79 demonstrate that it has a similar scale to the radio source (Batcheldor et al. 2006); allowing for projection effects, it is more likely that the radius we have used is conservative and the outflow region is significantly {\it larger} than we have estimated. Similarly, given that warm gas has been accelerated to large velocities and other compact radio sources show strong evidence for a relatively high density NLR (Holt et al. 2003, Holt 2005), it is unlikely that the electron density in the outflow can be substantially less than the measured density
in the narrow component ($430\pm50$~cm$^{-3}$) and the lower limit
($N \geq 10^2$ cm$^{-3}$) we have assumed. 

Given the fact that the quasar nucleus in PKS1549-79 is itself substantially obscured at optical wavelengths, it is less easy to dismiss the idea that a portion of the outflow region is highly extinguished at optical wavelengths so that the H$\beta$ luminosity is underestimated. However, it is unlikely that the H$\beta$ has been underestimated by more than a factor 50 because the blueshifted outflow component would then make a substantial contribution to the Pa$\alpha$ flux, which is not observed. 

Most plausibly, the warm gas outflow we have detected in [OIII] represents only a small fraction of the total mass in the outflow, with the remaining mass locked in hotter or cooler phases of the ISM that are difficult to detect at optical wavelengths. Recently, Morganti et al. (2003,2005a,b) have detected massive {\it neutral} outflows in several compact radio sources using observations of the 21cm line of neutral hydrogen. The mass outflow rates and energy fluxes contained within such neutral outflows are substantially larger than we have estimated for the warm gas outflow in PKS1549-79. It is notable that there is a hint of a blue wing to the HI~21cm line detected in PKS1549-79 (see Figures 11 and 12). If this can be confirmed with higher quality observations, it may be possible to make a more complete census of the warm and cool gas outflows in this object.
 
\subsection{HI spin temperature}

The exact value of the HI spin temperature ($T_{spin}$) is a major 
uncertainty in calculating the column density of neutral hydrogen using HI~21cm absorption
line observations. Although $T_{spin}$ is often assumed to have values consistent with
those estimated in  the Galaxy ($100 < T_{spin} < 1000$~K: Heiles \& Kulkarni 1988), higher values may be appropriate in the central regions around AGN because of excitation of the upper level of the 21cm hyperfine transition by
collisions, Ly$\alpha$ photons or 21cm continuum photons (see Bahcall \& Ekers 1969 for a discussion). 
The latter may be particularly important in a case like PKS1549-79 because the radio emission is beamed towards the HI absorbing cloud along our line of sight. However, to our knowledge there have so far
been no direct estimates of the HI 21cm spin temperature in the vicinity of luminous AGN.

PKS1549-79 provides an excellent opportunity to estimate the spin temperature directly because we have an independent estimate of the neutral column towards the quasar nucleus from the extinction
estimates derived from our quasar continuum modelling. The SED modelling gives 
$2.0 < E(B-V) < 4.0$ for the intrinsic line of sight reddening towards the quasar nucleus. Assuming the
standard Galactic dust to gas ratio ($E(B-V)/N_{HI} = 1.7\times10^{-22}$~mag cm$^{-2}$: Bohlin et al. 1978)
this translates into neutral HI columns in the range $1.2\times10^{22} < N_{HI} < 2.4\times10^{22}$~cm$^{-2}$. On the basis of our HI 21cm absorption line measurements, the neutral HI column is related to the spin temperature by:
$N_{HI} = (4.0\pm0.5)\times10^{18} T_{spin}$~cm$^{-2}$. Therefore, by combining the estimates of the neutral HI column derived from both the HI~21cm and the quasar SED measurements we find that the electron spin temperature
must fall in the range $3000 < T_{spin} <  6000$~K. These measurements provide clear evidence that the HI spin temperature in PKS1549-79 is higher than the typical Galactic value.

Implicit in our estimate of $T_{spin}$ is the assumption  that the column reddening the quasar is the same as that causing the HI 21cm absorption. Although the quasar nucleus can be considered point-like, while the radio emission is clearly spatially extended in our VLBI observations, our LBA observations show that the HI absorption covers the entire spatially-resolved radio source, including the flat spectrum core to the western jet of the source. Therefore it is entirely plausible that the HI 21cm and quasar reddening columns are associated with the same cloud. Another major assumption we have made is that the dust-to-gas ratio in the absorbing cloud has the Galactic value. This is called into question by recent observations that provide evidence for dust-to-gas ratios in the vicinity of AGN that are a factor $\sim$10 or more {\it lower} than the Galactic value (Maiolino et al. 2001). However, if such low dust-to-gas ratios also hold in PKS1549-79, the resulting spin temperature would be even larger than we have estimated above, further strengthening the evidence for large $T_{spin}$ in this object. In view of this
the lower limit we have derived for the spin temperature in PKS1549-79 ($T_{spin} > 3.0\times10^3$~K) is likely to be conservative.

\section{Conclusions and further work}

The new observations presented in this paper confirm that PKS1549-79 represents a luminous quasar observed at an early stage of its evolution. The quasar and jet activity have been triggered in a major merger which has also triggered substantial star formation. Accreting at close to the Eddington rate, the black hole in this system is growing rapidly and simultaneously driving relativistic jets. The growth phase of the black hole is substantially obscured at
optical wavelengths in this object, consistent with recent galaxy evolution
models (Hopkins et al. 2005). However, although the optical emission lines provide evidence for a high ionization, warm outflow associated with the  AGN, the warm outflow currently observed at optical wavelengths is not in itself sufficient to remove all the gas from the central regions of the galaxy. 

PKS1549-79 is a key object because it links together several classes of active galaxies, including radio galaxies, quasars, ULIRGs, NLS1 and AGN with high ionization absorption line systems. Moreover it shows many properties in common with the recently discovered population of high redshift AGN detected at sub-mm wavelengths (Alexander et al. 2005). The common thread is that all these classes of object are associated with the triggering and feedback effects of major galaxy mergers. Although much attention is paid to the links between star formation and activity in objects at high redshift, this study of PKS1549-79 serves to emphasise that some galaxies are evolving rapidly in the local universe, and that it is possible to use such objects to study the coevolution of black holes and galaxy bulges in depth.

Several further observational studies of PKS1549-79 would enhance our understanding of the relationship(s) between AGN activity and galaxy evolution. They include the following.
\begin{itemize}
\item [-] Near-IR measurements of the stellar bulge luminosity and velocity dispersion to improve estimates of the black hole mass.
\item [-] High quality optical spectroscopy observations of a wider range of emission line
diagnostics to measure the electron density accurately, and thereby improve our determination  of
the properties of the warm gas outflow. 
\item [-]Deep X-ray spectroscopy observations to refine estimates of the quasar bolometric luminosity and
absorbing HI column. Such observations would also allow  the links between PKS1549-79, NLS1 and sub-mm galaxies to be further investigated.
\item [-]Multi-epoch VLBI observations to search for superluminal motions in the jet, and accurately
determine the inclination of the jet to the line of sight.
\item [-]Deeper HI 21cm measurements and high resolution optical observations of the NaID line to investigate the importance of neutral gas outflows.  
\end{itemize}  

\subsection*{Acknowledgments} Based on observations collected at the European Southern Observatory, Chile
(ESO Programmes 69.B-0548(A), 71.B-0320(A) and 72.B-0792(A)) and the Anglo Australian Observatory, Siding Spring, Australia. KJI, JH and MB acknowledge financial support from PPARC. Edward King (CSIRO) provided original images made with the SHEVE  
network. We thank the anonymous referee for useful comments.
The LBA is part of the Australia Telescope, funded by the  
Commonwealth  of Australia for operation as a National Facility  
managed by CSIRO and the University of Tasmania. 

{}

\end{document}